\documentclass[11pt]{article}
\pdfoutput=1

\usepackage{epsfig}\parskip 5pt plus 1pt
\usepackage{amsmath}
\usepackage{amssymb}
\usepackage{amsfonts}
\usepackage{mathrsfs}
\usepackage{graphicx}
\usepackage{color}
\newcommand{\be}{\begin{equation}}
\newcommand{\ee}{\end{equation}}
\newcommand{\bea}{\begin{eqnarray}}
\newcommand{\eea}{\end{eqnarray}}

\selectfont

\begin{document}

\vspace{1cm}
\vspace*{-1.0truecm}
\begin{flushright}
UAB-FT-708\\
MPP-2012-96
\end{flushright}
\vspace{0.8truecm}

\begin{center}
\boldmath
{\Large\bf Postcards from oases in the desert:\\phenomenology of SUSY with intermediate scales}
\unboldmath

\end{center}
\vskip 0.4cm
\begin{center}
{\large Carla Biggio$^{\,a}$, Lorenzo Calibbi$^{\,b}$, Antonio Masiero$^{\,c}$, Sudhir K.~Vempati$^{\,d}$}
\vskip 0.4cm

{\footnotesize

$^a${\sl Institut de F\'\i sica d'Altes Energies, Universitat
Aut\`onoma de Barcelona, 08035 Bellaterra, Barcelona, Spain}

$^b${\sl Max-Planck-Institut f\"ur Physik (Werner-Heisenberg-Institut),
 F\"ohringer Ring 6, \\ D-80805 M\"unchen, Germany}

$^c${\sl Dipartimento di Fisica, Universit\`a di Padova, and INFN, Sezione
    di Padova, via F. Marzolo 8,\\ 35131 Padova, Italy}

$^d${\sl Centre for High Energy Physics, 
Indian Institute of Science, 
Bangalore 560 012, India}

}

\end{center}

\begin{abstract}
 The presence of new matter fields charged under the Standard Model
 gauge group at intermediate scales below the Grand Unification scale
 modifies the renormalization group evolution of the gauge
 couplings. This can in turn significantly change the running of the
 Minimal Supersymmetric Standard Model parameters, in particular the
 gaugino and the scalar masses. In the absence of new large Yukawa
 couplings we can parameterise all the intermediate scale models in
 terms of only two parameters controlling the size of the unified
 gauge coupling.  As a consequence of the modified running, the low
 energy spectrum can be strongly affected with interesting
 phenomenological consequences.  In particular, we show that scalar
 over gaugino mass ratios tend to increase and the regions of the
 parameter space with neutralino Dark Matter compatible with
 cosmological observations get drastically modified.  Moreover, we
 discuss some observables that can be used to test the intermediate
 scale physics at the LHC in a wide class of models.
\end{abstract}
\setcounter{footnote}{0}


\section{Introduction}

The apparent unification of the Standard Model (SM) gauge couplings
can be regarded as a major achievement of the Minimal Supersymmetric
Standard Model (MSSM).  Indeed, gauge couplings unification represents
the most convincing hint of a Grand Unified Theory (GUT) at very high
energy scales.  Typically, unification is achieved by assuming the
absence of new physics between the electroweak (EW) (or supersymmetric
(SUSY)) scale and the GUT scale, $M_{\rm GUT} \approx 10^{16}$ GeV.
In fact, the presence of fields charged under the SM gauge group at
intermediate scales below $M_{\rm GUT}$ modifies the renormalization
group (RG) evolution of the gauge couplings and in general spoils the
successful gauge coupling unification.  On the other hand, the
presence of ``oases'' of new physics in the ``big desert'' between
1~TeV and $M_{\rm GUT}$ is a natural prediction of many extensions of
the MSSM.  For instance, neutrino masses point towards a lepton number
breaking scale some orders of magnitude below $M_{\rm GUT}$.  The
fields associated with such new scale can be charged under the SM
gauge group, as in the case of the so-called type-II~\cite{typeII} or
type-III~\cite{typeIII,Abada:2007ux} seesaw models. Also, the
dynamical generation of the SM flavour hierarchy typically requires
heavy vectorlike quarks or Higgs fields as mediators of the flavour
symmetry breaking~\cite{FN} (for a recent discussion
see~\cite{High-Energy}).  Finally, intermediate scales are present in
models where the breaking of the GUT symmetry to the SM one is
achieved via intermediate steps, such as Pati-Salam models~\cite{PS}
or left-right symmetric models~\cite{LR}.

Gauge coupling unification can be maintained by appropriately choosing
the masses of the new fields or by embedding them in particular sets
(the simplest ones being complete multiplets of a GUT group, but more
general choices are also possible~\cite{Magic}).  Nevertheless, the
value of the unified coupling is modified by the presence of the new
fields.  This can in turn significantly change the RG running of the
MSSM parameters, in particular the gaugino and the scalar masses (if
the SUSY breaking occurs at scales higher than the intermediate
scale).  Therefore, one can expect a potentially observable impact of
the intermediate-scale physics on the low-energy SUSY spectrum and
phenomenology. Recently, the possible consequences of intermediate
scales have been discussed in a variety of specific
models~\cite{Buckley:2006nv}-\cite{Calibbi:2009wk}.

In this work we are going to discuss the phenomenological consequences
of the intermediate scale, due to the modified running of the MSSM
parameters. We assume that the SUSY breaking scale is equal or larger
than $M_{\rm GUT}$, such that the running of the SUSY breaking masses
is indeed affected by the presence of the intermediate scale.  Without
restricting to a specific model, we consider generic sets of new
matter fields (i.e.~chiral superfields) in vectorlike representations
of the SM gauge group, forming approximately degenerate multiplets of
a GUT group.  This allows us to highlight the common features and the
possible observable consequences of this kind of models, as it ensures
that gauge coupling unification is maintained independently of the
intermediate scale and $M_{\rm GUT}$ is the same as in the ordinary
MSSM.  These properties are in general not satisfied in the case of
multiple step breaking of the GUT symmetry to the SM, i.e.~in the presence
of new gauge bosons (vector superfields) below $M_{\rm GUT}$
(see however~\cite{Magic}).  Therefore, we assume the gauge group
to be $SU(3)\times SU(2)\times U(1)$ up to the GUT scale.  However, as
we will see, the effects we are going to discuss mainly depends on the
fact that, in the presence of new matter, the SM gauge couplings unify at
a common value that is larger than in the ``big desert'' scenario. As
a consequence, we expect that our findings qualitatively occur in
broader classes of models, whenever the latter feature is realised.

The rest of the paper is organised as follows: in
  section~\ref{sec:running} we discuss the main effect, i.e. the
  modification of the running of the MSSM parameters in the presence of
  new fields at intermediate scales. 
  In section~\ref{sec:LHC} we discuss a few
  observables that can be used to test the
  intermediate scale at the LHC, while in sections~\ref{sec:DM} and
  \ref{sec:pdecay} we analyse the consequences for the neutralino
  relic density and the proton decay,
  respectively. Finally, in section~\ref{sec:conclu} we conclude. 
   Details on analytical solutions of the one loop running 
   and a discussion on the impact of two loop RGEs are given 
   in the two appendices.

\section{MSSM running with an intermediate scale}
\label{sec:running}

\begin{figure}[!t]
\centering
\includegraphics[width=0.4\textwidth]{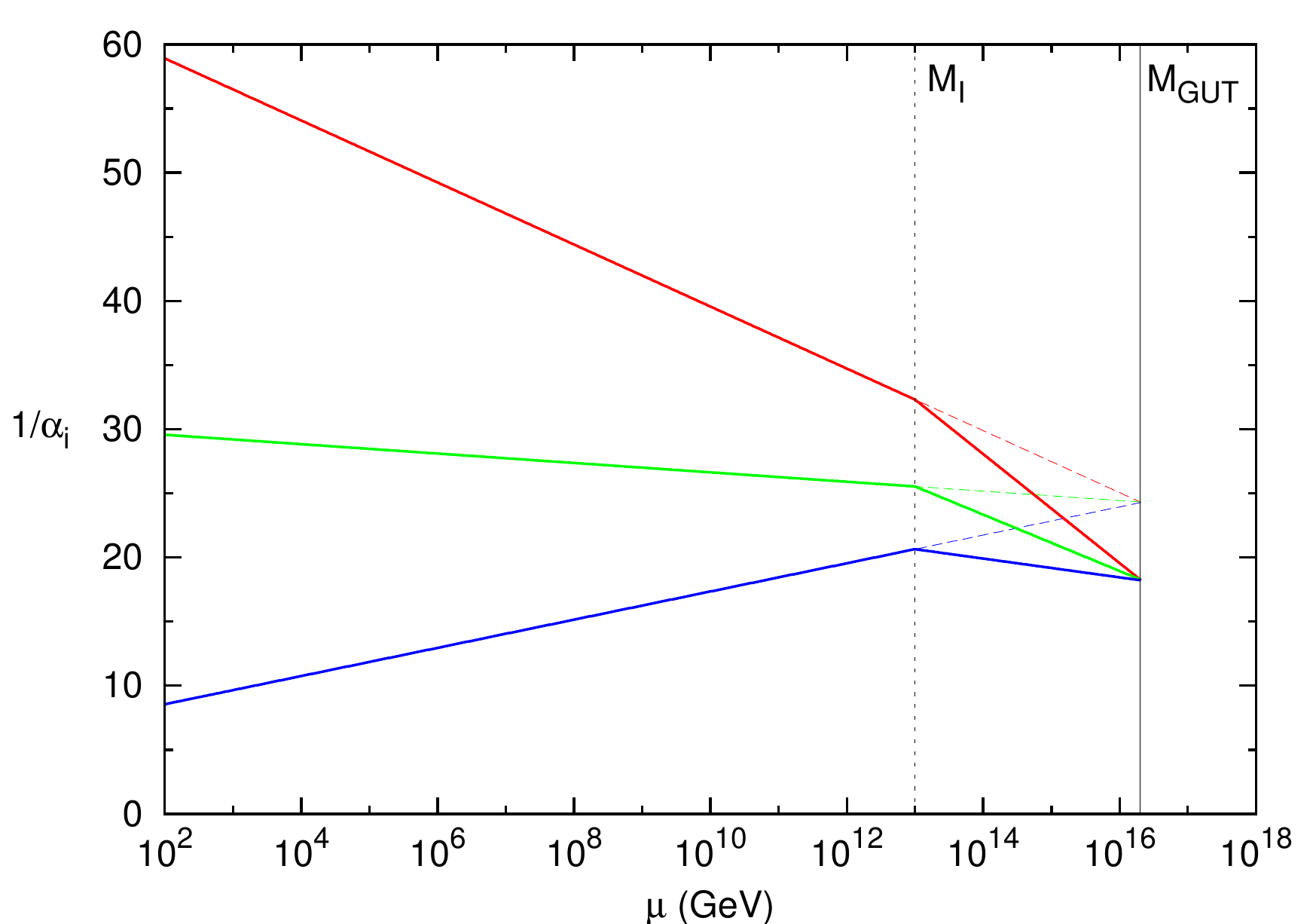}
\hspace{0.5cm}
\includegraphics[width=0.4\textwidth]{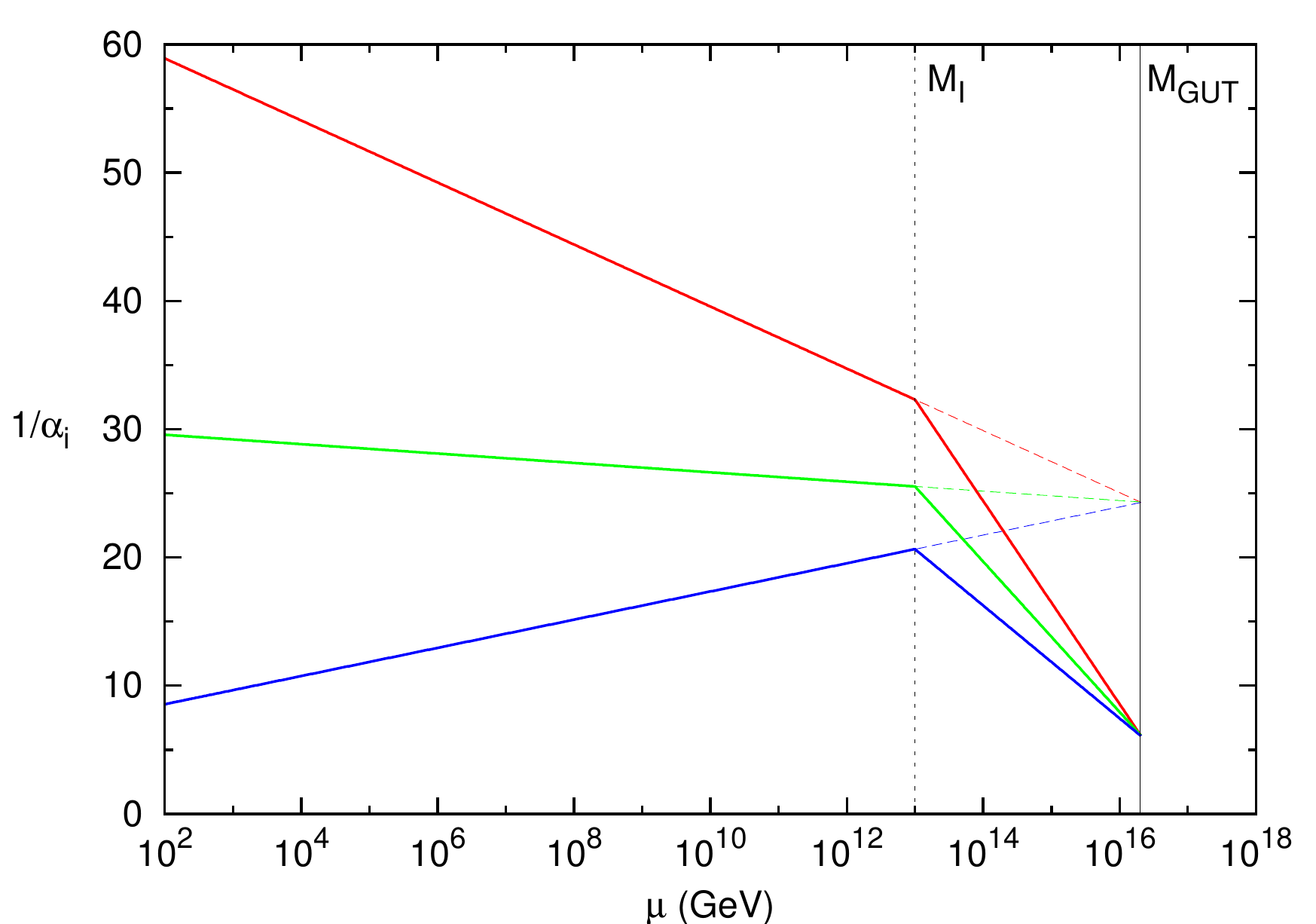}
\caption{Modification of the gauge couplings running in the presence
  of matter at the intermediate scale $M_I = 10^{13}$ GeV. In the left
  panel we take $\Delta b = 5$ (corresponding e.g.~to a single $\bf
  24$ representation of $SU(5)$), in the right panel we take $\Delta b
  = 15$ (e.g.~$3\times {\bf 24}$).}
\label{fig:gauge-coupl}
\end{figure}

Our starting assumption is the presence of a set of chiral superfields
in complete vectorlike representations of SU(5) at an intermediate
scale $M_I < M_{\rm GUT}$.  This choice does not spoil the successful,
one loop, gauge coupling unification of the MSSM, as the running of
the three couplings gets deflected in the same way. In other words,
the successful prediction for $\alpha_3 (M_Z)$ (and $M_{\rm GUT}$) is
not modified.\footnote{This conclusion holds under the assumption that
  there are no large mass splittings among the fields in the SU(5)
  multiplets.}  However, it is well known that the fields at $M_I$
make the running above this scale ``stronger'' and the gauge couplings
finally unify at a value $\alpha_U$ larger than in the MSSM.  This
effect can be seen by solving the one loop RGEs:
\begin{equation}
 \frac{1}{\alpha_U} = \frac{1}{\alpha_i (M_Z)} 
- \frac{b^{SM}_i}{2 \pi} \ln \frac{M_S}{M_Z} 
- \frac{b^0_i}{2 \pi} \ln \frac{M_{\rm GUT}}{M_S} 
- \frac{\Delta b}{2 \pi} \ln \frac{M_{\rm GUT}}{M_I} \equiv \frac{1}{\alpha^0_U} 
- \frac{\Delta b}{2 \pi} \ln \frac{M_{\rm GUT}}{M_I} \,, 
\label{eq:alpharun}
\end{equation}
where $\alpha^0_U$ is the unified coupling in the MSSM without
intermediate scale ($\alpha^0_U\simeq 1/25$), $b_i^{SM} =
(41/10,-19/6,-7)$ and $b^0_i = (33/5,1,-3)$ are respectively the SM
and MSSM $\beta$-function coefficients for $\alpha_i$ ($i=1,2,3$),
$\Delta b$ is the universal contribution of the additional fields at
$M_I$ and $M_S$ is the typical low-energy SUSY scale.  $\Delta b$ is
given by the sum of the Dynkin indexes of the SU(5) representations of
the fields at $M_I$.\footnote{For example in the SU(5) embedding of
  type-II seesaw~\cite{Rossi:2002zb}, the new fields are in a {$\bf
    15+\overline{15}$} representation, which gives $\Delta b = 7$,
  while in SUSY type-III seesaw~\cite{Perez:2007rm,Biggio:2010me,Esteves:2010ff} 
  each copy of {$\bf 24$} contributes with $\Delta b = 5$. We
  remind that a copy of {$\bf 5+\overline{5}$} corresponds to $\Delta
  b =1$.}  From Eq.~(\ref{eq:alpharun}), we see that, since $\Delta b
\geq 0$ for chiral superfields, the unified coupling $\alpha_U$ is in
general larger than the MSSM one, $\alpha_U \geq
\alpha^0_U$.\footnote{For vector superfields $\Delta b$
    would be negative, implying a reduction of the value of $\alpha_U$
    and a consequent modification of all the effects discussed
    here. However, since new gauge groups are usually accompanied by
    new chiral superfields, as long as the net effect is an increment
    of $\alpha_U$, the results discussed here will qualitatively
    hold.}  This effect is exemplified in Fig.~\ref{fig:gauge-coupl},
for $M_I = 10^{13}$ GeV and $\Delta b = 5,~15$.  The dashed lines
represent the ordinary MSSM running.

\begin{figure}[t]
\centering
\includegraphics[width=0.4\textwidth]{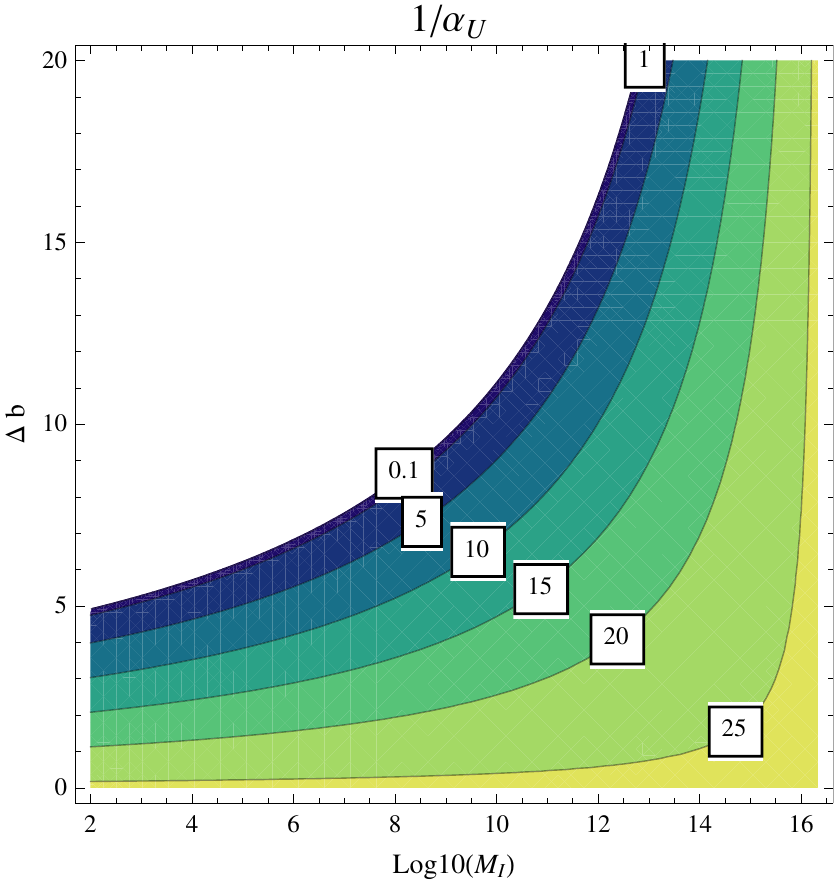}
\hspace{0.4cm}
\includegraphics[width=0.42\textwidth]{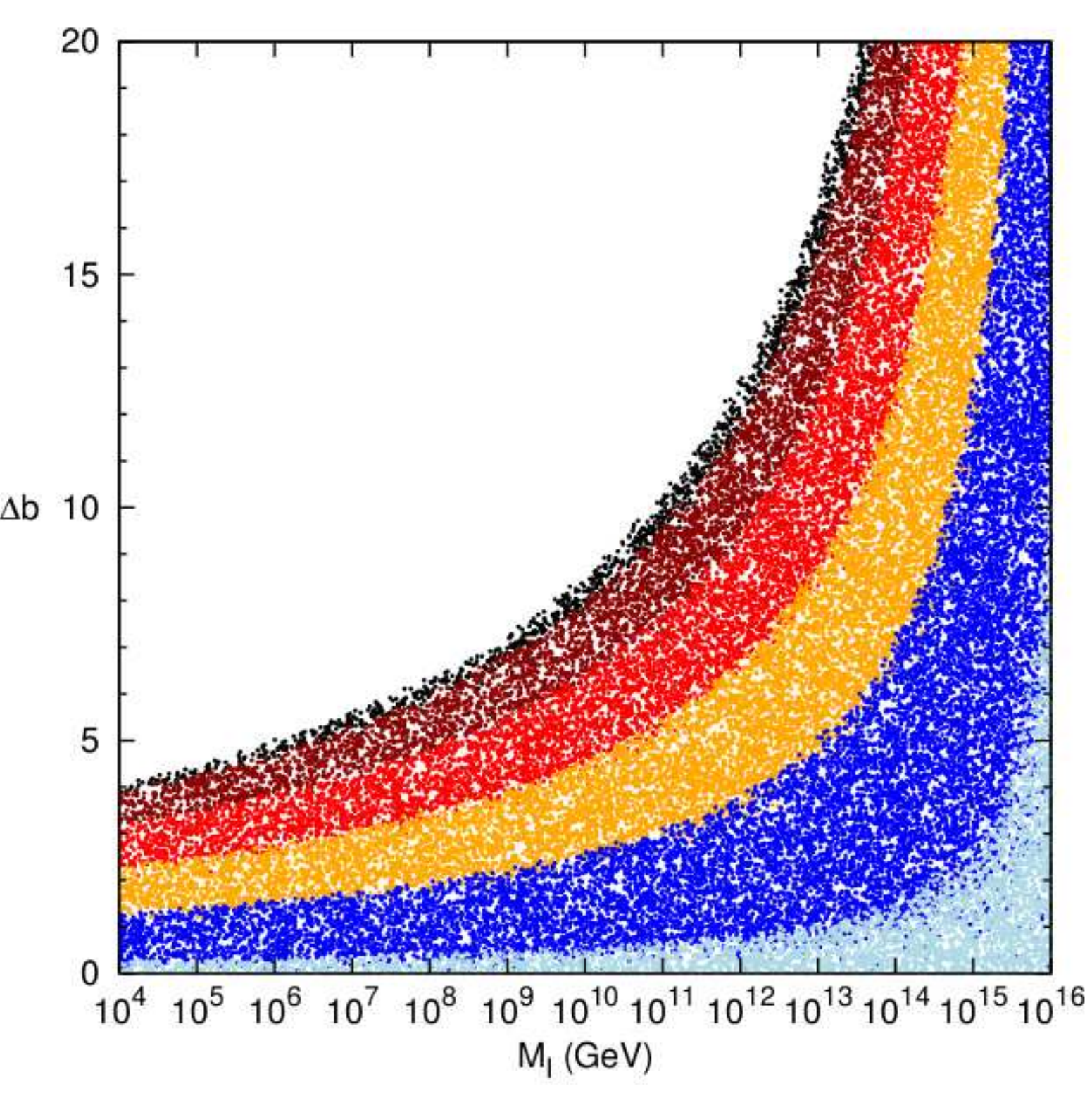}
\caption{Contours on the plane $M_I$-$\Delta b$ of the inverse of the
  unified gauge coupling $1/\alpha_U$ (considering one loop RGEs in
  the left panel, two loops in the right one).  The white region is
  excluded from the requirement of perturbativity of the couplings up
  to the GUT scale. In the right panel light blue corresponds to
  $1/\alpha_U>25$, blue to $>20$, yellow to $>15$, red to $>10$, brown
  to $>5$ and black to $>0$. We refer to Appendix~B for details on the
  meaning of $\Delta b$ in the two loop case.}
\label{fig:perturbativity}
\end{figure}

Clearly, for a given $\Delta b$, Eq.~(\ref{eq:alpharun}) will give a
lower bound on the scale $M_I$ by requiring perturbativity of the
gauge couplings up to the GUT scale. We can see the perturbativity
bound in Fig.~\ref{fig:perturbativity}, where contours for
$1/\alpha_U$ are plotted on the $M_I$-$\Delta b$
plane.\footnote{Notice however that $\Delta b$ is a discrete
  quantity.}  The white area in the plot is excluded since it
corresponds to $1/\alpha_U < 0$, i.e. to a Landau pole below the GUT
scale.  Looking at the left panel, where results obtained using
Eq.~(\ref{eq:alpharun}) are shown, we see for instance that with
$\Delta b = 15$ (e.g. corresponding to $3\times {\bf 24}$), the
intermediate scale is constrained to be $M_I \gtrsim 10^{12}$ GeV,
while in case of the type-II seesaw ($\Delta b = 7$) we have $M_I
\gtrsim 10^{7}$ GeV.  On the other hand, it is remarkable that $M_I$
can be as low as the TeV scale, provided that $\Delta b \lesssim
5$.\footnote{This result is well known in the context of
  gauge-mediated SUSY breaking, see e.g.~\cite{GMSB}.}
We thus observe that already with this simple requirement, we can
exclude a large part of the parameter space. 

The figure in the left panel was derived using Eq.~(\ref{eq:alpharun})
but does not qualitatively change considering two loops RGEs, as can
be seen in the right panel of
Fig.~\ref{fig:perturbativity}. Notice, in particular,
  that the region far from the Landau pole is practically unaltered,
  while large modifications appear for $\alpha_U \gtrsim 0.2$. It is evident
  from this plot that constraints derived at one loop
are then conservative. Details on two loops RGEs will be given in Appendix B.

As we will see in the following, the main effects we are going to
discuss are linked to the larger values of $\alpha_U$ induced by the
intermediate scale physics and therefore can be conveniently
illustrated in terms of two additional parameters only, $M_I$ and
$\Delta b$.

\subsection{Running of gaugino masses}

\begin{figure}[!h]
\centering
\includegraphics[width=0.4\textwidth]{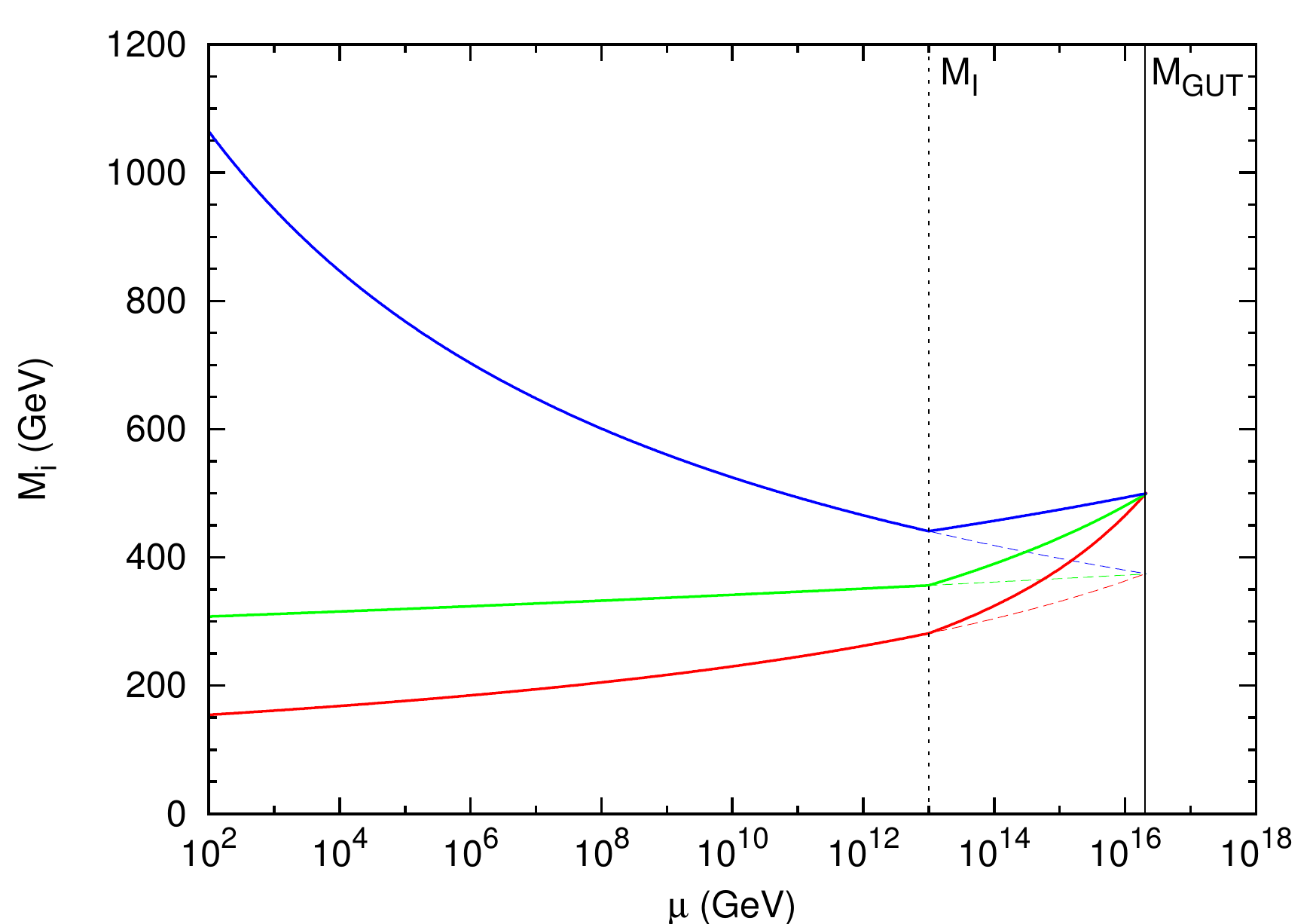}
\hspace{0.5cm}
\includegraphics[width=0.4\textwidth]{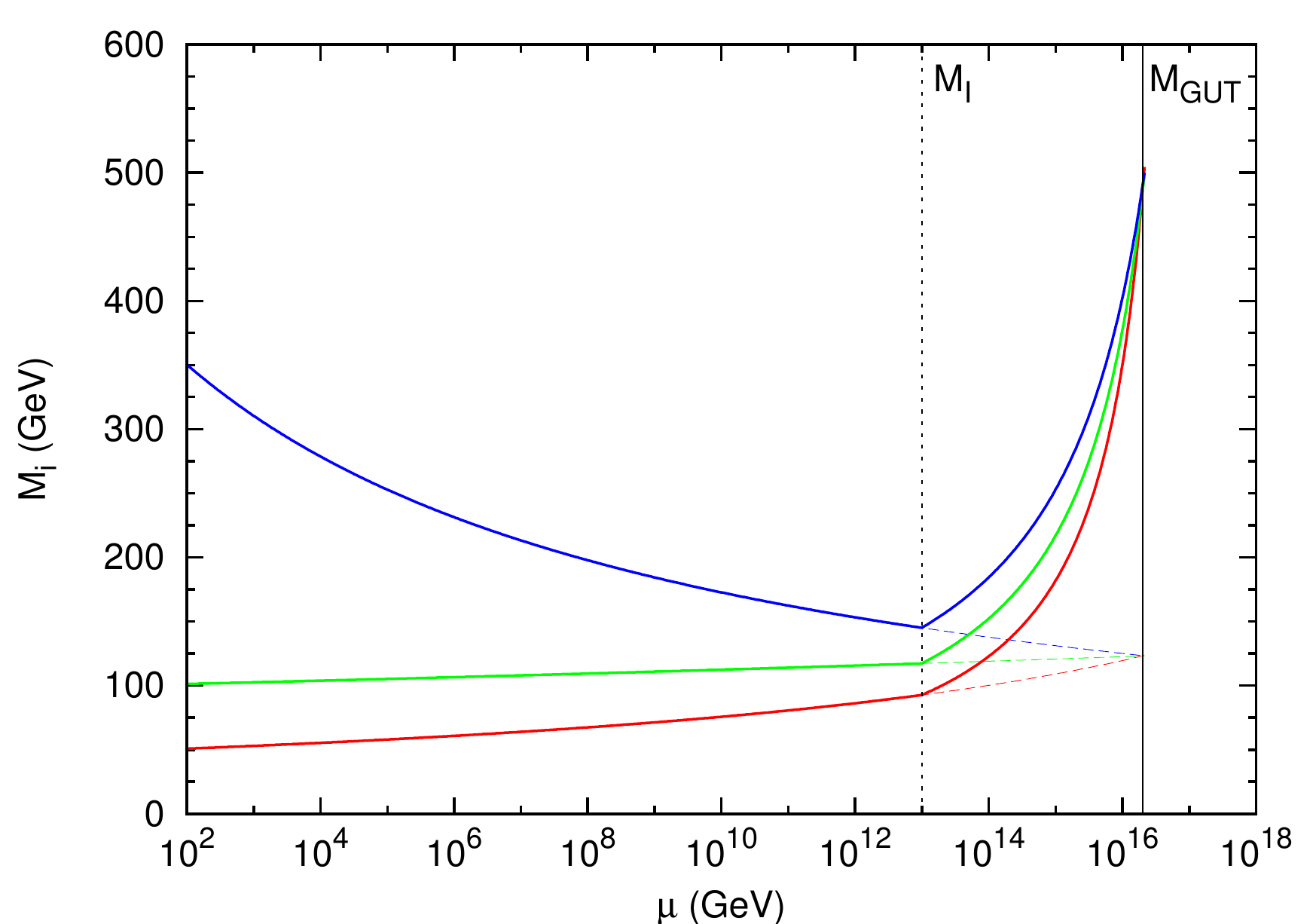}
\caption{Running of gaugino masses for $\Delta b = 5$ (left), $\Delta
  b = 15$ (right).  Solid lines correspond to the running with new
  matter at the intermediate scale $M_I = 10^{13}$ GeV, while dashed
  lines represent the ordinary MSSM running, corresponding to the same
  values for $M_i$ at low energy.}
\label{fig:gaugino}
\end{figure}

Let us now move to consider the effect of intermediate-scale physics
on the gaugino mass running.  As we know, the $\beta$-functions of the
gaugino masses are related to those of the corresponding gauge
coupling, hence the modification of the running of the gauge couplings
above $M_I$ will affect the running of the gaugino mass parameters,
$M_i$ ($i=1,2,3$), as well. This effect can be easily related to the
increase of the unified gauge coupling $\alpha_U$. We can see this
from the usual one loop relation among gaugino masses and gauge
couplings, which is not modified in our scenario:
\begin{equation}
 M_i(\mu) = M_i(M_{\rm GUT})\,\frac{\alpha_i(\mu)}{\alpha_U} \,,
\label{eq:gauginos}
\end{equation}
where $\mu$ is the renormalization scale. Obviously, the low-energy
gaugino mass ratios ($M_1:M_2:M_3 \approx 1:2:6$ in the case of
gaugino mass unification) are then not modified by the presence of the
intermediate scale.  However, for given initial values $M_i(M_{\rm
  GUT})$, the low-energy gaugino masses result smaller than in the
MSSM, since $\alpha_U$ is larger.  This effect is depicted in
Fig.~\ref{fig:gaugino} for the same choices of $M_I$ and $\Delta b$ of
Fig.~\ref{fig:gauge-coupl}, assuming gaugino mass unification with
$M_i(M_{\rm GUT})=M_{1/2}$.
Above the scale $M_I$, gaugino masses can have a very strong running,
while below the usual MSSM evolution is clearly recovered.  The dashed
lines in the plots represent the ordinary MSSM running, giving the
same gaugino mass spectrum at low energy.  It is clear from
Eq.~(\ref{eq:gauginos}) that, in order to obtain the same gaugino
masses at low energy, the starting unified mass $M^0_{1/2}$ in absence
of the intermediate scale must be smaller than $M_{1/2}$:
\begin{equation}
M^0_{1/2} = M_{1/2} \, \frac{\alpha_U^0}{\alpha_U} = M_{1/2} \left(1 - \alpha_U^0 \frac{\Delta b}{2 \pi} \ln \frac{M_{\rm GUT}}{M_{I}}\right)\, .
\end{equation}
At this stage, this might be seen as a trivial rescaling of $M_{1/2}$,
i.e.~a larger value at $M_{\rm GUT}$ than in the MSSM is required to
reproduce a given gaugino spectrum in the presence of the intermediate
scale. However, this affects non-trivially the running of other MSSM
parameters, in particular scalar masses, as we are going to discuss in
the following.

\subsection{Running of scalar masses}
In order to see the effect on the evolution of the scalar masses, let
us recall the form of the one loop RGEs for the soft SUSY breaking
mass terms. Denoting sfermion and Higgs fields as $\phi$, we
schematically have:
\begin{equation}
 \frac{d}{dt} m^2_{\phi} = - \frac{2}{\pi} \sum_i C_i(r_\phi) \alpha_i(t) |M_i(t)|^2 + \frac{1}{16 \pi^2} (Y^2 m^2_{\phi^\prime}
 + A^2)\,,
\label{eq:scalarrun}
\end{equation}
where we can choose $t=\ln(\mu/M_{\rm GUT})$ and $C_i(r_\phi)$ is the
quadratic Casimir of the representation $r_\phi$ of the field $\phi$.
$Y^2$ generically denotes Yukawa couplings (with $m^2_{\phi^\prime}$
we indicate that soft masses of different scalars can appear) and
similarly $A^2$ schematically refers to contributions proportional to
the A-terms. From Eq.~(\ref{eq:scalarrun}), we can see the well known
behaviour in the scalar masses evolution: in the running from $M_{\rm
  GUT}$ to low energies the gauge part of the $\beta$-function
($\propto \alpha_i(t) |M_i(t)|^2$) tends to increase the scalar mass
$m^2_{\phi}$, while the terms proportional to the Yukawa and the
trilinear couplings have the opposite effect and tend to decrease
it. This latter effect can be however sizeable only if large third
generation Yukawas and A-terms are involved (such as in the case of
the stop and $H_u$ masses, where these terms are proportional to
$y_t^2$).  Therefore for what concerns 1st and 2nd generation sfermion
masses, we can consider only the gauge term in the $\beta$-functions
and obtain simple analytical solutions of the one loop RGEs (see
Appendix A).\footnote{Later on, when we will discuss the effects of
  the intermediate scale on the electroweak symmetry breaking (EWSB)
  or on third generation sfermion masses, for which the Yukawa
  contribution cannot be neglected, we will solve numerically the full
  set of two loops RGEs, cf.~Appendix B.}
\begin{figure}[!t]
\begin{center}
\includegraphics[width=0.3\textwidth]{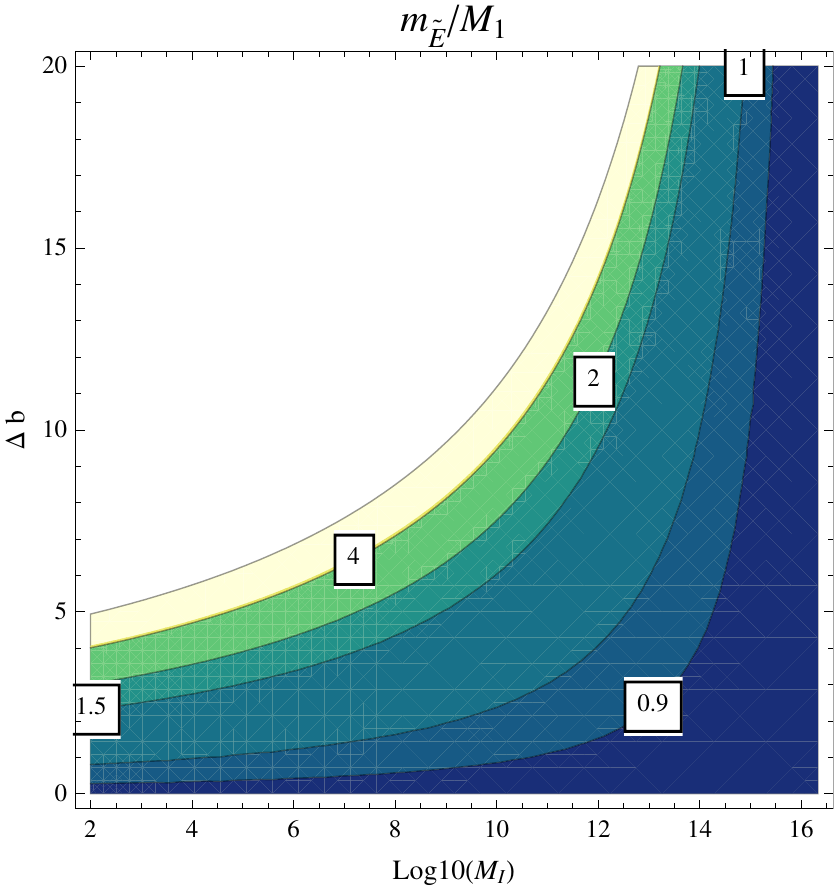}
\hspace{0.2cm}
\includegraphics[width=0.3\textwidth]{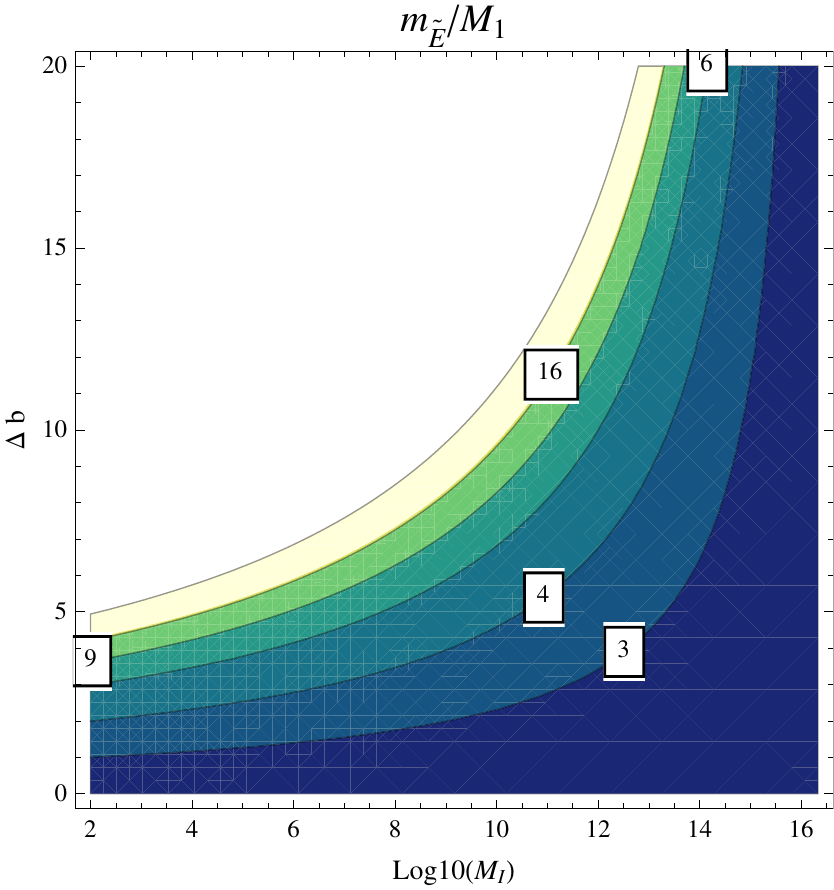}
\hspace{0.2cm}
\includegraphics[width=0.3\textwidth]{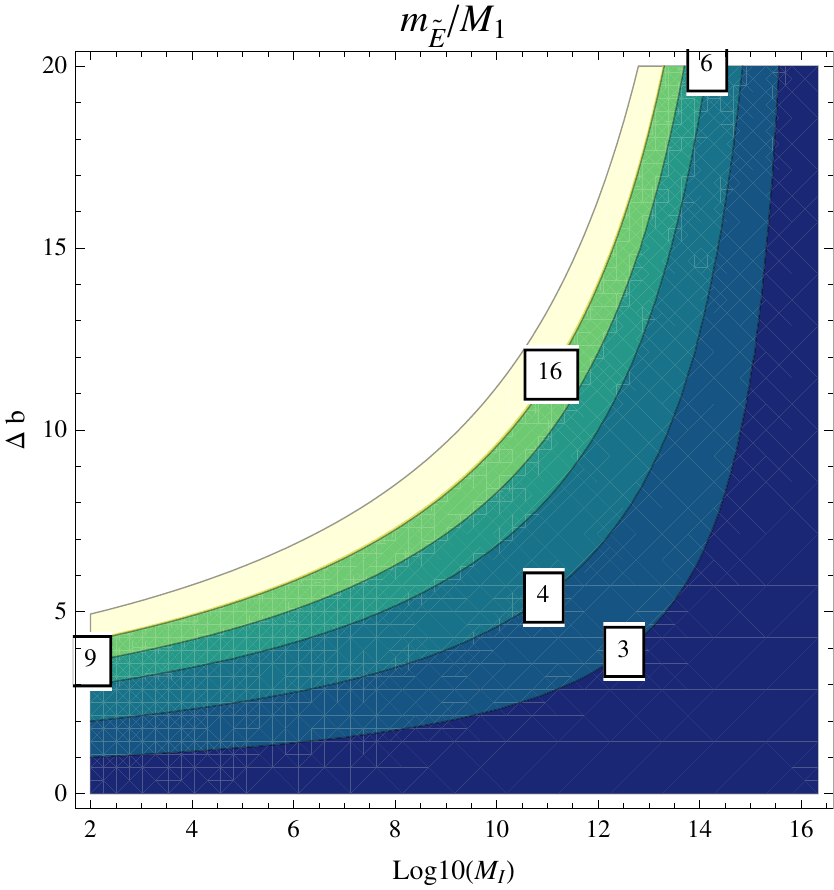}
\caption{Ratio of the 1st or 2nd generation RH sleptons over the bino
  mass $M_1$ for $m_{\tilde E}/M_{1}=0,\, 1,\, 2$ at $M_{\rm GUT}$.}
\label{fig:ratio-l}
\end{center}
\end{figure}
\begin{figure}[!t]
\centering
\includegraphics[width=0.3\textwidth]{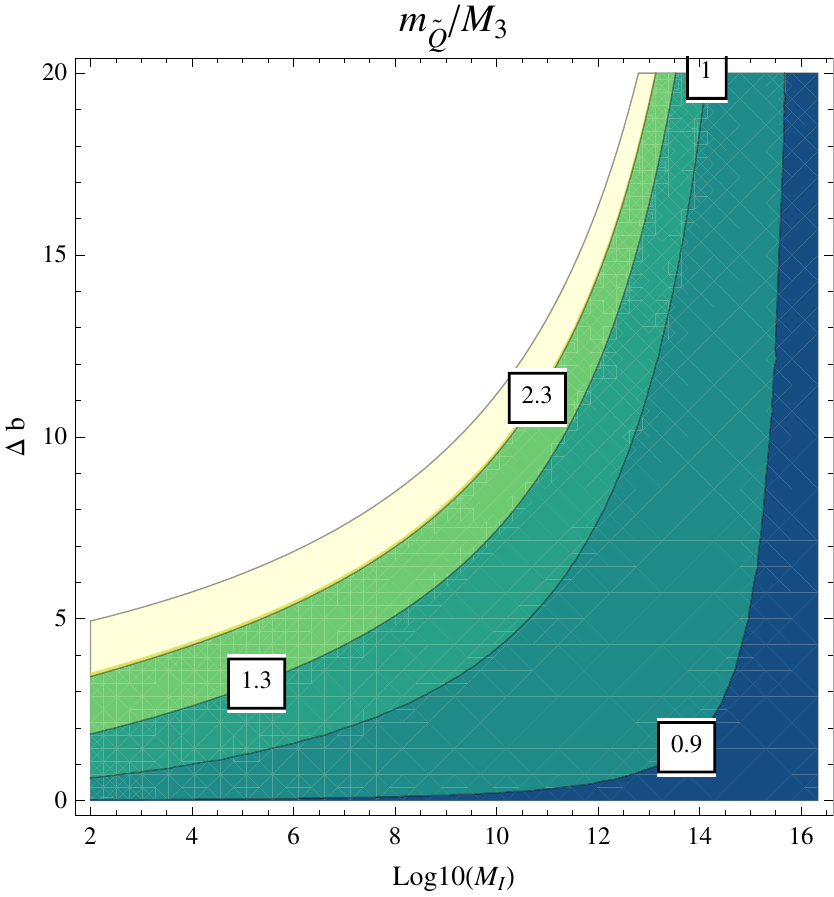}
\hspace{0.2cm}
\includegraphics[width=0.3\textwidth]{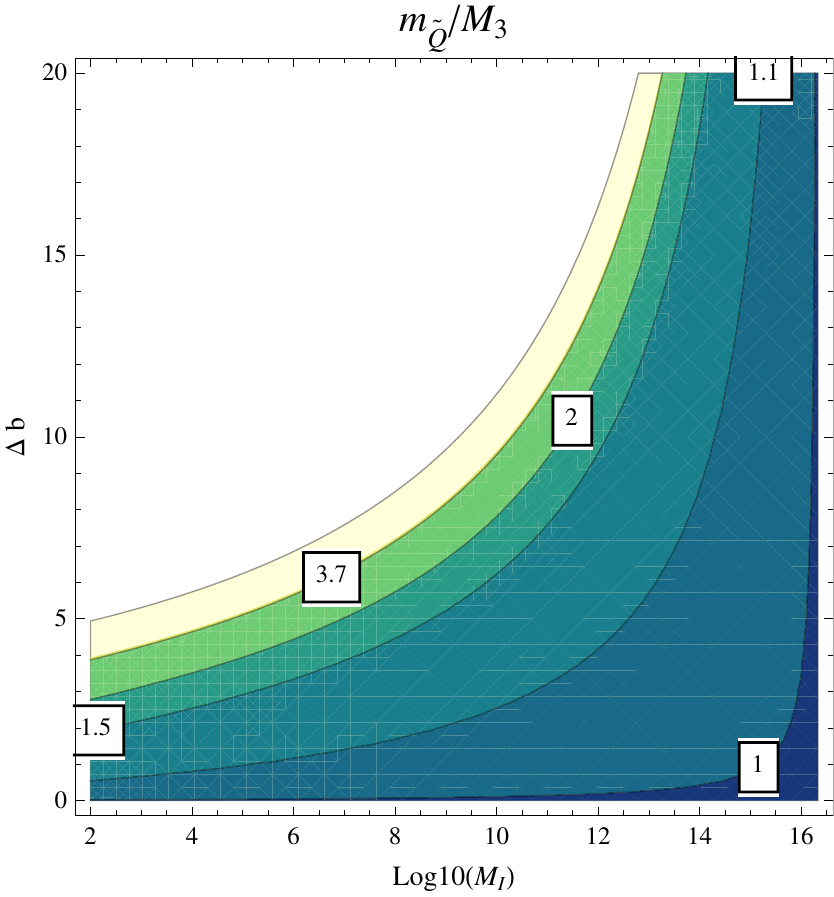}
\hspace{0.2cm}
\includegraphics[width=0.3\textwidth]{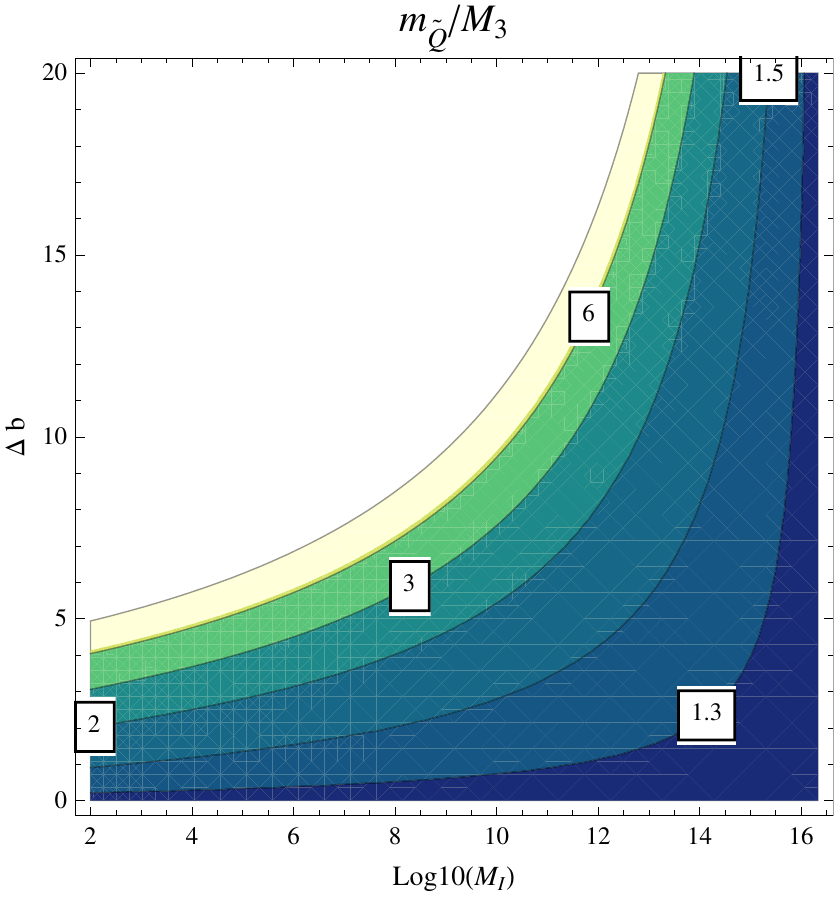}
\caption{Ratio of the 1st or 2nd generation LH squarks over the gluino
  mass $M_3$ for $m_{\tilde Q}/M_{3}=0,\, 1,\, 2$ at $M_{\rm GUT}$.}
\label{fig:ratio-q}
\end{figure}

How is the running of scalar masses affected by the intermediate scale
fields? As we have seen in Figs.~\ref{fig:gauge-coupl} and
\ref{fig:gaugino}, $\alpha_i$ and $M_i$ run to larger values above
$M_I$.  Starting with the same scalar and gaugino masses at $M_{\rm
  GUT}$ and running down to the EW scale, in the presence of the
intermediate scales, scalar masses will grow less than in the MSSM
case, because of the fast decrease of gaugino masses shown in
Fig.~\ref{fig:gaugino}.  But while considering only gaugino masses the
MSSM spectrum could be recovered just by rescaling the GUT values
$M_i(M_{\rm GUT})$, this is not any longer true if scalars are also
taken into account. In other words, the intermediate scale has
low-energy consequences in form of a distortion of the SUSY spectrum.
In fact, for the same values of low-energy gaugino masses as in the
MSSM, scalar masses feel an enhancement of the gauge part of the
$\beta$-function in Eq.~(\ref{eq:scalarrun}) in the first stage of the
running, between $M_{\rm GUT}$ and $M_I$, due to larger couplings and
heavier gauginos at high energy.  This means that the scalar masses
run to larger values, or, more precisely, the presence of
intermediate-scale fields tends to increase the scalar over gaugino
mass ratios, $m_\phi / M_i$.
Notice that here, in order to restrict our analysis to model
independent effects, we are assuming no large (i.e.~$\mathcal{O}(1)$)
Yukawa couplings among the MSSM fields and the new matter. In that
case the scalar masses would receive a further negative contribution,
as shown in Eq.~\ref{eq:scalarrun}. Such an assumption is also
motivated by the protection from new large flavour violating effects:
for instance in type II or III seesaw models the new Yukawas are
typically constrained to values that have a negligible impact on the
SUSY spectrum ($\lesssim(0.1)$) by the bounds on lepton flavour
violating processes~\cite{Biggio:2010me,Esteves:2010ff,Rossi:2002zb}.

The effect sketched above can be seen by looking at the RGE of such
mass ratios.  In particular, let us consider the ratio of the
right-handed (RH) selectron and Bino masses, ${m^2_{\tilde
    E}}/{M_1^2}$:
\begin{equation}
\frac{d}{dt} \left(\frac{m^2_{\tilde{E}}}{M_1^2} \right)=
- \frac{6}{5\pi} \alpha_1 
- \left(\frac{m^2_{\tilde{E}}}{M_1^2} \right) \frac{b_1}{\pi} \alpha_1
\, .
\end{equation}
Given the minus sign in the $\beta$-function, the ratio increases
in the running from high to low scale and, in the presence of
intermediate scale physics, it increases more, due to a larger gauge
coupling in the high-energy part of the running.  This can be also
shown by means of the one loop formulae of the Appendix A.  From
Eqs.~(\ref{eq:scalarmasses1}, \ref{eq:scalarmasses2}), we get:
\begin{align}
 \frac{m^2_{\tilde E}}{M_1^2}(M_{S}) =&\frac{m^2_{\tilde E}(M_{\rm GUT})}{M_1^2(M_{S})}+ \frac{6}{5 b^0_1} \left[\frac{b^0_1}{b^0_1+\Delta b} 
\frac{\alpha_U^2}{\alpha_1^2(M_S)} + \frac{\Delta b}{b^0_1+\Delta b} \frac{\alpha_1^2(M_I)}{\alpha_1^2(M_S)}- 1\right]\, . 
\end{align}
For a given low-energy value for $M_1(M_{S})$ and a given high-energy
starting value $m_{\tilde E}(M_{\rm GUT})$, the ratio above tends to
grow in the presence of the intermediate scale (i.e.~increasing $\Delta b$
and/or decreasing $M_I$), since ${\alpha_U^2}/{\alpha_1^2(M_S)}$
strongly grows, according to Eq.~(\ref{fig:gauge-coupl}).

The effect exemplified above is general for all scalar masses.
This is the main point of our discussion and it is represented in
Figs.~{\ref{fig:ratio-l}}-{\ref{fig:ratio-q}}, where the low-energy
ratios of the 1st or 2nd generation RH sleptons, $m_{\tilde E}$, over
the Bino mass $M_1$ and the 1st or 2nd generation left-handed (LH)
squarks, $m_{\tilde Q}$, over the gluino mass $M_3$ are plotted for
different GUT values of, respectively, $m_{\tilde E}/M_{1}$ and
$m_{\tilde Q}/M_{3}$.  Again, we made use of the analytical formulae
given in the Appendix A.  As in Fig.~\ref{fig:gauge-coupl}, the white
area corresponds to a Landau pole occurring below the GUT scale.  The
fact that low-energy mass ratios increase by introducing new matter at
the intermediate scale (i.e.~with increasing $\alpha_U$) leads to
potentially observable consequences at the LHC and affects DM
phenomenology, as we will discuss in the following sections.

\subsection{Higgs soft masses and EWSB}\label{sec:ewsb}

Apart from scalar and gaugino masses, the presence of intermediate
scale clearly affects the running of the Higgs mass parameters
$m^2_{H_u}$ and $m^2_{H_d}$ too, and thus affects the EWSB and the
higgsino mass $\mu$. Once correct EWSB is imposed, $\mu$ is given (at
tree level) by the well-known expression:
\begin{equation}
 |\mu|^2 = - m^2_{H_u} - \frac{M_Z^2}{2} + \mathcal{O}(m^2_{H_{u,d}}/(\tan\beta)^2)\, .
\label{eq:mur}
\end{equation}
The higgsino mass $\mu$ enters the neutralino mass matrix and it is
thus crucial to determine the composition of the lightest neutralino
and whether it can be a good dark matter (DM) candidate.
\begin{figure}[t]
\begin{center}
\includegraphics[width=0.55\textwidth]{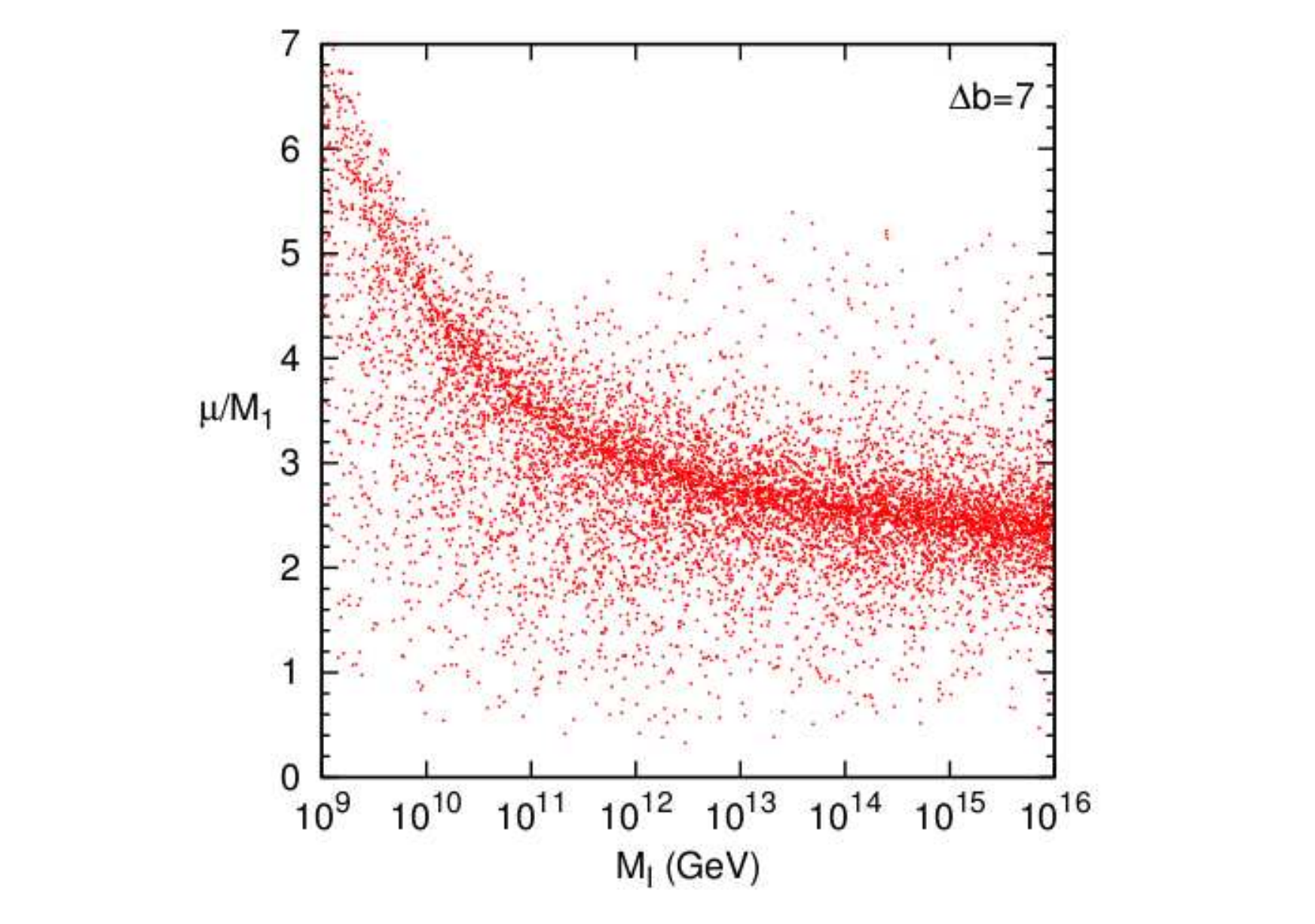}
\caption{Ratio $\mu/M_1$ vs.~$M_I$ for a numerical scan of the
  parameter space with $\Delta b$=7.}
\label{fig:ratio-muM1}
\end{center}
\end{figure}

In the usual MSSM $m_{H_u}$ runs to negative values at low energy due
to the terms in the $\beta-$function proportional to $y_t^2$, the top
Yukawa coupling.  In spite of the fact that $y_t$ at high energy is
somewhat smaller when an intermediate scale is present,\footnote{The
  reason is that the terms in the $y_t$ $\beta$-function proportional
  to the gauge couplings increase above $M_I$.} the increase of the
scalar over gaugino mass ratio turns out to be the dominant effect.
Since stop masses enter the terms $\propto y_t^2$ in the RGEs, the
net effect is that the ratio $m_{H_u}/M_i$ tends to run to even more
negative values when new physics at intermediate scale is present and
thus $\mu/M_i$ gets increased (cf.~Eq.~(\ref{eq:mur})).  This effect
in shown Fig.~\ref{fig:ratio-muM1} for the illustrative case of
$\Delta b$=7 and universal boundary conditions at the GUT scale.

The clear tendency shown in the figure generically tells us that: (i)
intermediate scale models tend to worsen the fine-tuning problem of
the MSSM, (ii) the lightest neutralino $\tilde \chi_1^0$ tends to be
more and more $\tilde B$-like.  However, independently of how strong
is the effect of the intermediate scale, this does not rule
out configurations in the parameter space that give $\mu\approx M_1$,
i.e.~with low fine-tuning and a neutralino relic density in agreement
with WMAP observations, due to the sizable higgsino component of
$\tilde \chi_1^0$~\cite{focuspoint}. Furthermore, non-universal Higgs
boundary conditions at the high scale can affect this result.

\section{LHC observables}
\label{sec:LHC}

As we discussed above, intermediate scale physics can leave a clear
imprint on the low-energy SUSY spectrum. The question is whether this
can be observed at the LHC.  In order to isolate the effect of the
intermediate scale, we need observables that are as much independent
as possible of high-scale scalar and gaugino masses.  For instance,
even if the low-energy scalar over gaugino mass ratios grow in the
presence of intermediate scale fields, as shown in
Figs.~\ref{fig:ratio-l} and \ref{fig:ratio-q}, there is also a strong
dependence on the high-energy initial conditions, hence it seems hard
to disentangle the two effects.  However, even at this stage, we can
see that SUSY searches at the LHC can be affected by the intermediate
scale.  Let us consider, for instance, the first panel of
Fig.~\ref{fig:ratio-q}, that corresponds to the case of vanishing
squark masses at the GUT scale. If we do not allow $m^2_{\tilde
  Q}(M_{\rm GUT})<0$, such configuration clearly gives the least
possible ratio $(m_{\tilde Q}/M_3)^{\rm min}$ at low energy .  While
in the ordinary MSSM $(m_{\tilde Q}/M_3)^{\rm min} \approx 1$, we see
that the intermediate scale can easily push the minimum ratio to
values larger than 2.  This means that the configuration $M_{\tilde
  g}\approx m_{\tilde Q}$ that gives the highest sensitivity in the
LHC SUSY searches (see e.g.~\cite{Baer}) would not be theoretically
accessible and, independently of the starting values for the soft
masses at high energy, only the case $M_{\tilde g} < m_{\tilde Q}$ (or
even $M_{\tilde g} \ll m_{\tilde Q}$) would be possible. On the other
hand, observing the case $M_{\tilde g}\approx m_{\tilde Q}$ at the LHC
would give an upper bound to $\alpha_U$ and thus strongly constrain
the $\Delta b$ and $M_I$ parameters.

In the following we will describe other quantities that are to large
extent model-independent and can be used to constrain the presence of
new physics at intermediate scales.

\subsection{Mass Invariants}

\begin{figure}[t]
\centering
\includegraphics[width=0.3\textwidth]{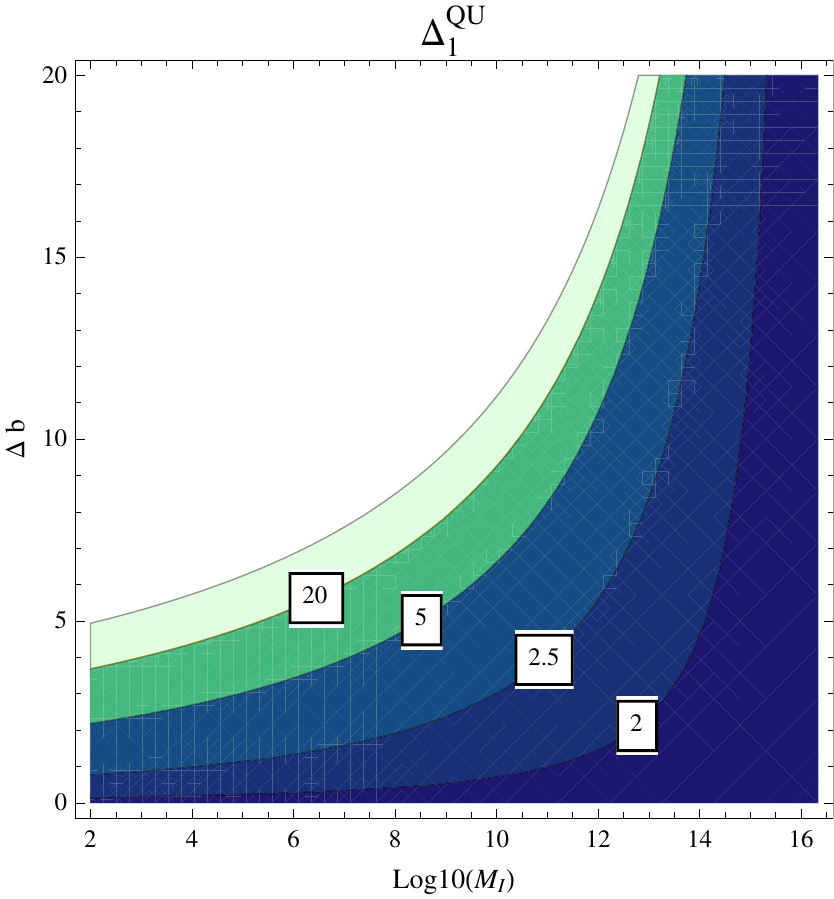}
\hspace{0.2cm}
\includegraphics[width=0.3\textwidth]{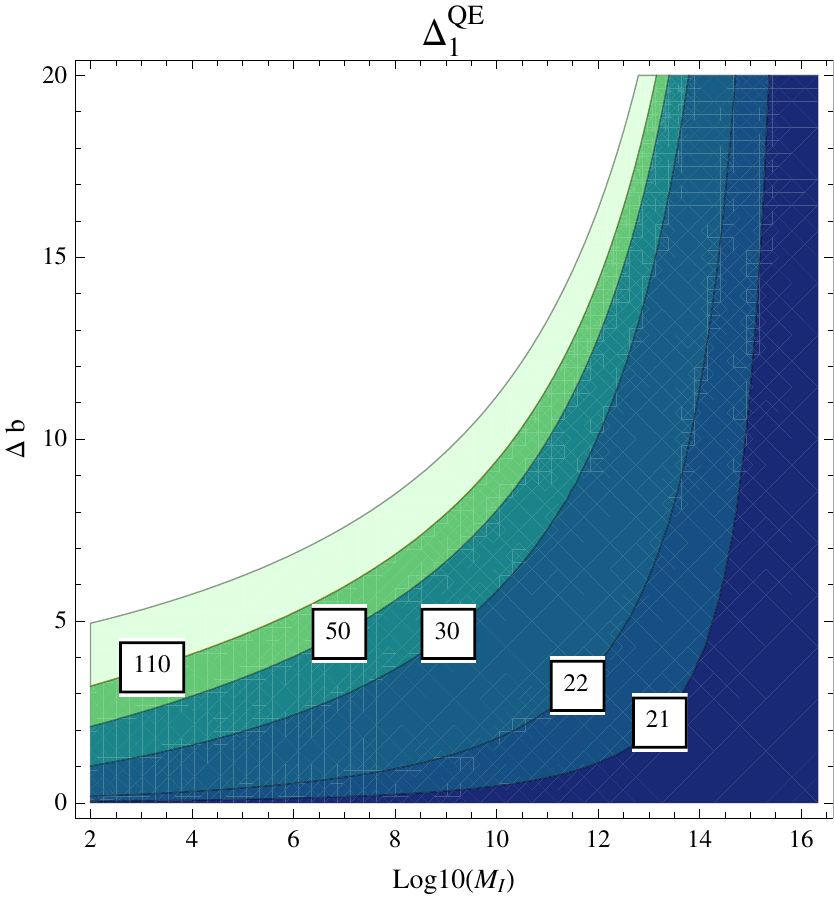}
\hspace{0.2cm}
\includegraphics[width=0.3\textwidth]{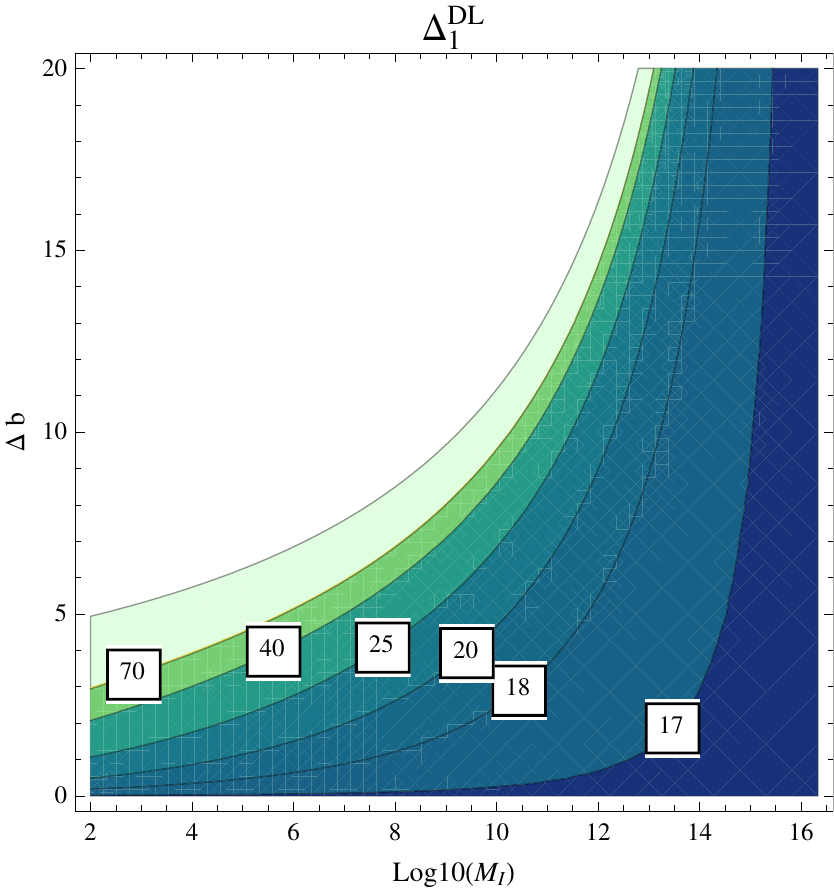}
\caption{Contours for the mass invariants $\Delta^{QU}_1$, $\Delta^{QE}_1$, $\Delta^{DL}_1$, defined in Eq.~(\ref{eq:SU5inv}), for M$_{S}=1$~TeV.}
\label{fig:BMinv}
\end{figure}

If we assume gaugino mass unification, the gaugino and first generations sfermion masses can be written (at one loop) in the form:
\begin{eqnarray}
M_i(M_S) &=& A_i(M_S,\Delta b,M_I)\, M_{1/2} \\
m^ 2_{\tilde{f}}(M_S) &=& m_{\tilde f}^2(M_{\rm GUT}) + B_{\tilde{f}}(M_S,\Delta b,M_I)\, M_{1/2}^2 \,, 
\end{eqnarray}
where the coefficients $A_i$ and $B_{\tilde{f}}$, as functions of $\Delta b$ and $M_I$, 
can be read in Eqs.~(\ref{eq:gaugino-sol}) and (\ref{eq:scalarmasses1}, \ref{eq:scalarmasses2}). 

It is clear that in the combination
\begin{equation}
\Delta^{ff^\prime}_i\equiv \frac{m^ 2_{\tilde{f}}-m^ 2_{\tilde{f'}}}{M_i^2}
\label{eq:massinv}
\end{equation}
the explicit dependence on the GUT-scale parameters drops, if
$m_{\tilde f^\prime}^2(M_{\rm GUT})=m_{\tilde f}^2(M_{\rm GUT})\equiv
m_0^2$ as in CMSSM-like scenarios or, more in general, in the case of
GUT-symmetric initial conditions (a well-motivated assumption in our
setup, as we are requiring unification).  Notice, however, that $A_i$
and $B_{\tilde{f}}$ do not depend on $\Delta b$ and $M_I$ only, but
logarithmically on the SUSY mass-scale $M_S$ as well.  This induces a
residual dependence of the parameters $\Delta^{ff^\prime}_i$ on the
initial conditions $m_0$ and $M_{1/2}$ that can be relevant, as we are
going to show in the following.

In Ref.~\cite{Buckley:2006nv}, it has been pointed out that mass
invariants of the kind of Eq.~(\ref{eq:massinv}) are very sensitive to
intermediate scale fields and can be thus useful to discriminate among
different SUSY seesaw models (see
also~\cite{DeRomeri:2011ie,Esteves:2011gk}).  Here we want to
generalise that result and study these invariants for generic values
of $M_I$-$\Delta b$.  As in Ref.~\cite{Buckley:2006nv}, we consider
the SU(5)-inspired combinations:
\begin{equation}
\Delta^{QU}_1 \equiv\frac{m^2_{\tilde{Q}}-m^2_{\tilde{U}}}{M_1^2},\quad\quad \Delta^{QE}_1 \equiv\frac{m^2_{\tilde{Q}}-m^2_{\tilde{E}}}{M_1^2},
\quad\quad \Delta^{DL}_1 \equiv\frac{m^2_{\tilde{D}}-m^ 2_{\tilde{L}}}{M_1^2}.
\label{eq:SU5inv}
\end{equation}
Contours for these quantities on the $M_I-\Delta b$ plane are shown in
Fig.~\ref{fig:BMinv} (taking $M_S = 1$~TeV).  As we can see, the
invariants rapidly grow for increasing $\alpha_U$. This is a further
consequence of the effect described in the previous section:
$m^2_{\tilde{Q}}/M_1^2$ increases with $\alpha_U$ and does it more
than $m^2_{\tilde{U}}/M_1^2$ that does not feel the contribution of
SU(2) gauginos, thus $\Delta^{Q U}_1$ grows too. An analogous effect
occurs for the other invariants.  Measurements of the SUSY spectrum at
the LHC can be then potentially used to derive precise information on
the nature of the new physics possibly present at intermediate
scales.\footnote{We remind that we are considering only new chiral
  superfields at $M_I$ and thus we assume only the SM gauge group up
  to GUT scale. In the presence of new gauge groups, the contribution
  to the invariants of the new vector superfields can be negative, see
  e.g.~\cite{DeRomeri:2011ie,Esteves:2011gk}, so that cancellations
  might occur.}  If the invariants of Eqs.~(\ref{eq:SU5inv}) will be
measured to differ significantly from the MSSM values
($\Delta^{QU}_1\approx 2$, $\Delta^{QE}_1\approx 21$,
$\Delta^{DL}_1\approx 17$ for $M_S= 1$ TeV), this might be a hint of
new physics below the GUT scale (that can provide information on the
scale and nature of it), or simply a signal that soft masses are not
universal at the GUT scale.  In principle, it should be still possible
to identify the first case by reconstructing more than one invariant,
since the three invariants exhibit robust correlations, as
Fig.~\ref{fig:BMinv} shows.

As in the case of $\alpha_U$, the above effect is strengthened at two loops, especially in the vicinity of the Landau pole. The typical correction is however $\lesssim$ 10\% (see Appendix B) 
and hence the plots of Fig.~\ref{fig:BMinv} still give a good estimate of the impact of the intermediate scale on these quantities.

As we mentioned above, the numerical values of the invariants still depend on the SUSY scale. For instance, in the range $M_Z \le M_S \le 3$ TeV, we observe a variation of $\Delta^{ff^\prime}_i$ up to $60\%$, which of course might spoil any attempt to constrain intermediate scale with them. 
Even if it is true that once the sparticle masses will be measured, also the SUSY scale can be set, we look for mass invariants which are more independent on the SUSY scale.

\begin{figure}[t]
\centering
\includegraphics[width=0.35\textwidth]{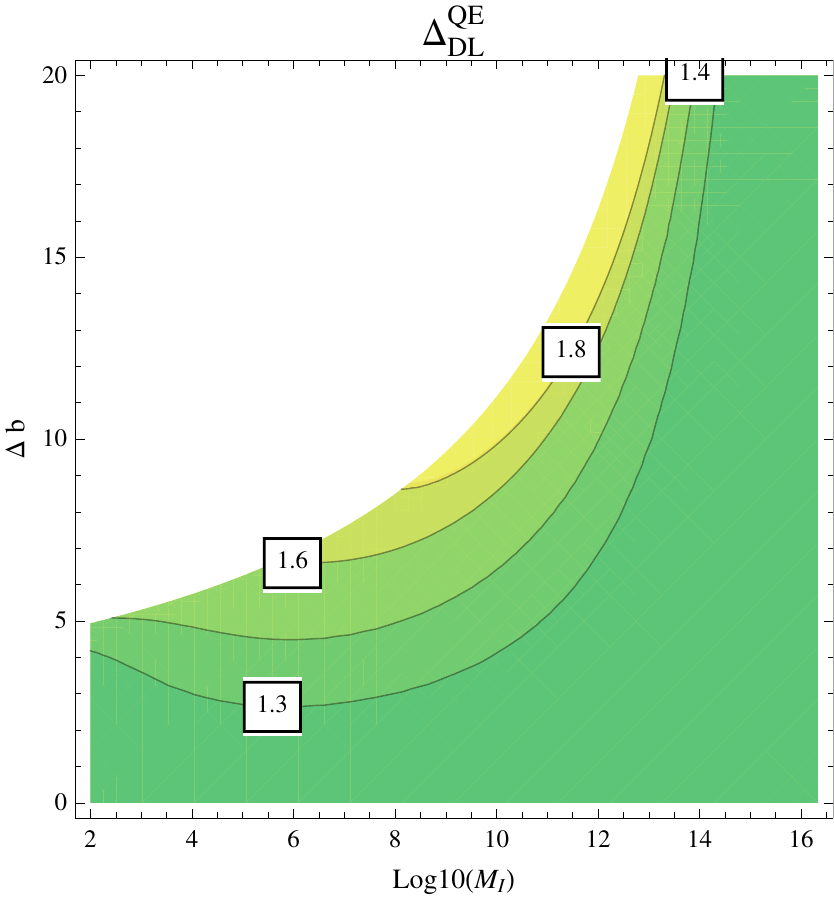}
\hspace{0.5cm}
\includegraphics[width=0.35\textwidth]{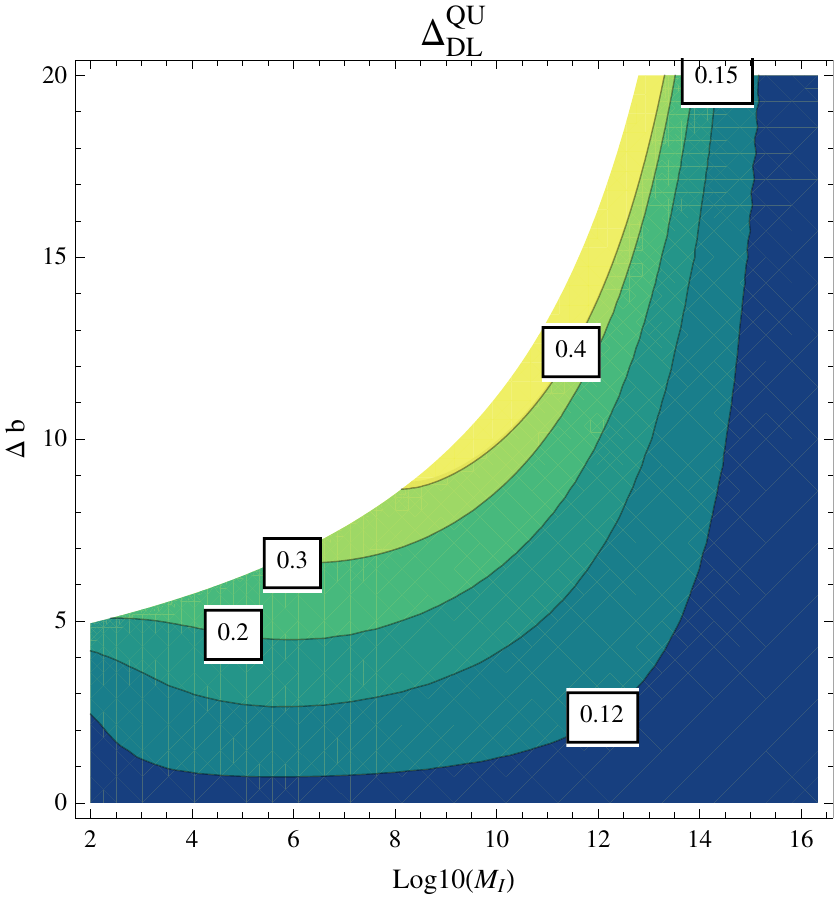}
\caption{Contours for the mass invariants $\Delta^{QE}_{DL}$, $\Delta^{QU}_{DL}$, defined in Eq.~(\ref{eq:CBinv}), for M$_{S}=1$~TeV.}
\label{fig:BCinv}
\end{figure}

We consider ratios of the above invariants, i.e.~simply ratios of scalar mass differences:
\begin{equation}
\Delta^{f_1 f_2}_{f_3 f_4} \equiv \frac{m^ 2_{\tilde{f}_1}-m^ 2_{\tilde{f}_2}}{m^ 2_{\tilde{f}_3}-m^ 2_{\tilde{f}_4}}\, .
   \label{eq:CBinv}
\end{equation}
Compared to the quantities in Eq.~(\ref{eq:SU5inv}), these ones have a much milder dependence on $M_{S}$. This is true in particular for 
$\Delta^{QE}_{D L} = (m^2_{\tilde Q}-m^2_{\tilde E})/(m^2_{\tilde D
}-m^2_{\tilde L})$, where the variation in the range $M_Z \le M_S \le
3$ TeV is just of a few percent. 
Countour plots for these invariants are shown in Fig.~\ref{fig:BCinv}.
Besides using the analytical one loop expressions, we numerically computed the invariants including two loops contributions. The results are shown in Appendix B. The 
effect of the intermediate scale is again stronger than at one loop and deviates from Fig.~\ref{fig:BCinv} quite sizeably close to the Landau pole.

We have shown that mass invariants can in principle give information
and strong constraints on $\Delta b$ and $M_I$.  They might even
exclude, if observed to be close to the CMSSM predictions, the
presence of fields charged under the SM gauge group at scales below
$M_{\rm GUT}$.  However, it is difficult to say whether all the SUSY
masses entering these invariants can be measured with sufficient
precision at the LHC.  Clearly this question depends on the actual
mass-scale of the SUSY particles, as well as on some features of the
spectrum, and is beyond the purposes of the present discussion.  Let
us only notice that, if the slepton masses can be reconstructed from
cascade decays of heavier particles (see the next section), the
invariants involving squark and slepton mass differences should not be
difficult to obtain, given the typical hierarchy between coloured and
uncoloured sfermions.  Moreover, Figs.~\ref{fig:BMinv} and
\ref{fig:BCinv} show that, even in presence of uncertainties as large
as 10\%, the invariants can provide very useful information on the
intermediate scale and discriminate among different scenarios.  It
might be much more difficult to resolve experimentally squark mass
differences like $m^2_{\tilde Q}-m^2_{\tilde U}$. 
However, we notice that intermediate scale
physics might help in this sense. In fact, the relative mass-splitting
$(m_{\tilde Q}-m_{\tilde U})/m_{\tilde Q}$ (that can at most be
$\approx 5$\% in the CMSSM) increases with $\alpha_U$ too and can
reach values larger than 10\%.

\subsection{Kinematic Edges in Cascade Decays}

\begin{figure}[h]
\begin{center}
\includegraphics[width=0.45\textwidth]{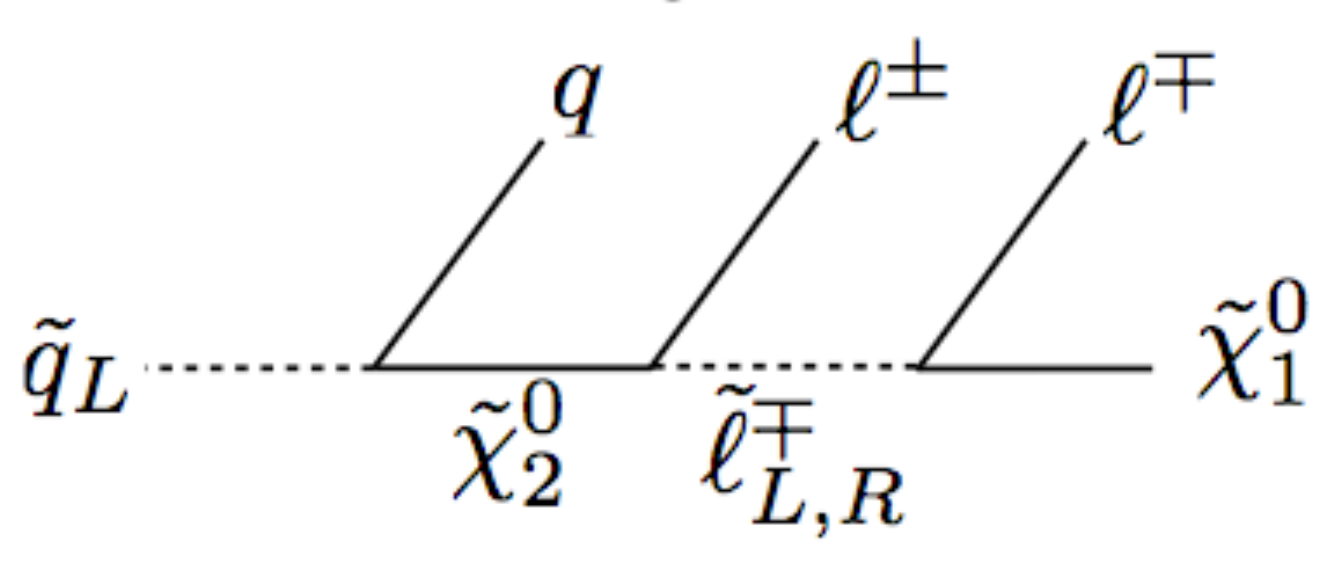}
\end{center}
\caption{An example of a cascade decay.}
\label{fig:cascade}
\end{figure}

Let us now investigate whether the distortion of the spectrum can be studied by means of kinematic observables potentially measurable at the LHC experiments. 
We consider the typical cascade decay depicted in
Fig.~\ref{fig:cascade}. 
The intermediate particles are real if the following condition is fulfilled:
\begin{equation}
m_{\tilde{Q}} > m_{{\tilde\chi}^0_2} > m_{\tilde{\ell}_{L,R}} > m_{{\tilde\chi}^0_1} \, .
\label{eq:cascade}
\end{equation}
Consequently, the invariant-mass distributions of the outgoing SM
particles (jets and isolated leptons) exhibit sharp kinematic
end-points~\cite{Bachacou:1999zb}. Notice that, depending on the
spectrum, zero, one or two sharp edges can be present. Indeed, if for
example both $m_{\tilde{\ell}_{L}}$ and $m_{\tilde{\ell}_{R}}$ satisfy the
above inequality, then two edges could be observed, while only one
will be there if only one of the two --typically $m_{\tilde{l}_{R}}$--
satisfies it. The position of the end-points of the distributions can
be expressed as a function of the SUSY masses:
\begin{eqnarray}
m_{\ell\ell}^{\rm max} &=& \sqrt{\frac{(m^2_{{\tilde\chi}^0_2} - m^2_{\tilde{\ell}}) (m^2_{\tilde{\ell}} - m^2_{{\tilde\chi}^0_1})}{m^2_{\tilde{\ell}}}}\\
m_{\ell j}^{\rm max} &=& \sqrt{\frac{(m_{\tilde{q}}^2-m^2_{{\tilde\chi}^0_2})(m^2_{{\tilde\chi}^0_2}-m_{\tilde{\ell}}^2)}{m^2_{{\tilde\chi}^0_2}}} \\
m_{\ell\ell j}^{\rm max} &=& \sqrt{\frac{(m_{\tilde{q}}^2-m^2_{{\tilde\chi}^0_2})(m^2_{{\tilde\chi}^0_2}-m^2_{{\tilde\chi}^0_1})}{m^2_{{\tilde\chi}^0_2}}} \, ,
\end{eqnarray}
and can therefore be used to reconstruct the SUSY
spectrum~\cite{Bachacou:1999zb}. As any combination of SUSY masses,
these observables are also modified in the presence of
the intermediate scale.
\begin{figure}[t]
\centering
\includegraphics[height=0.3\textwidth]{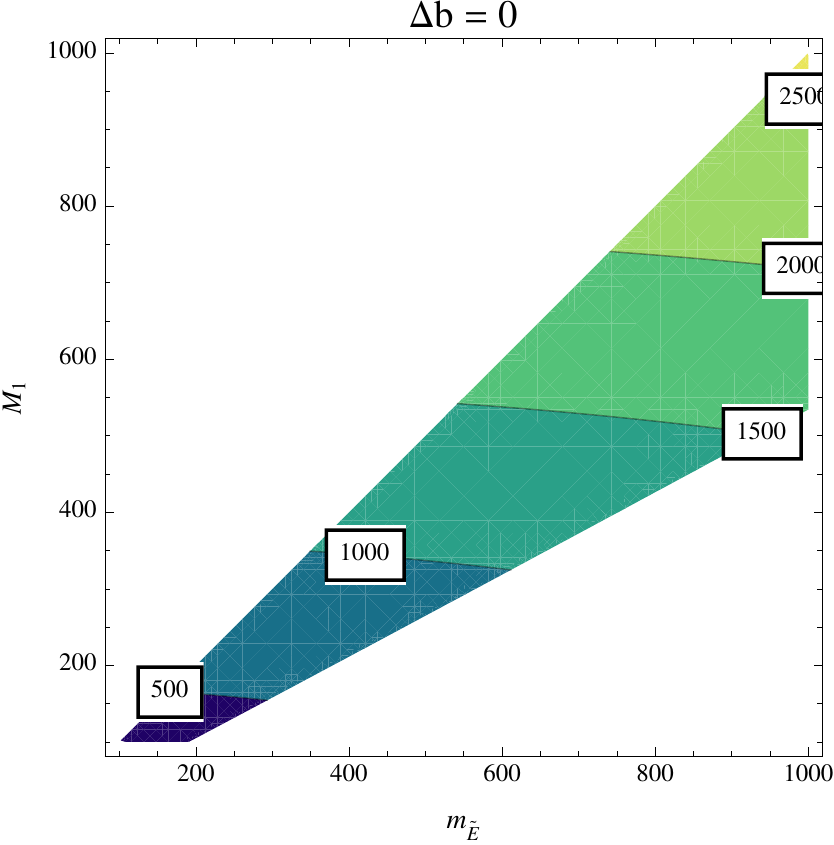}
\hspace{0.2cm}
\includegraphics[height=0.3\textwidth]{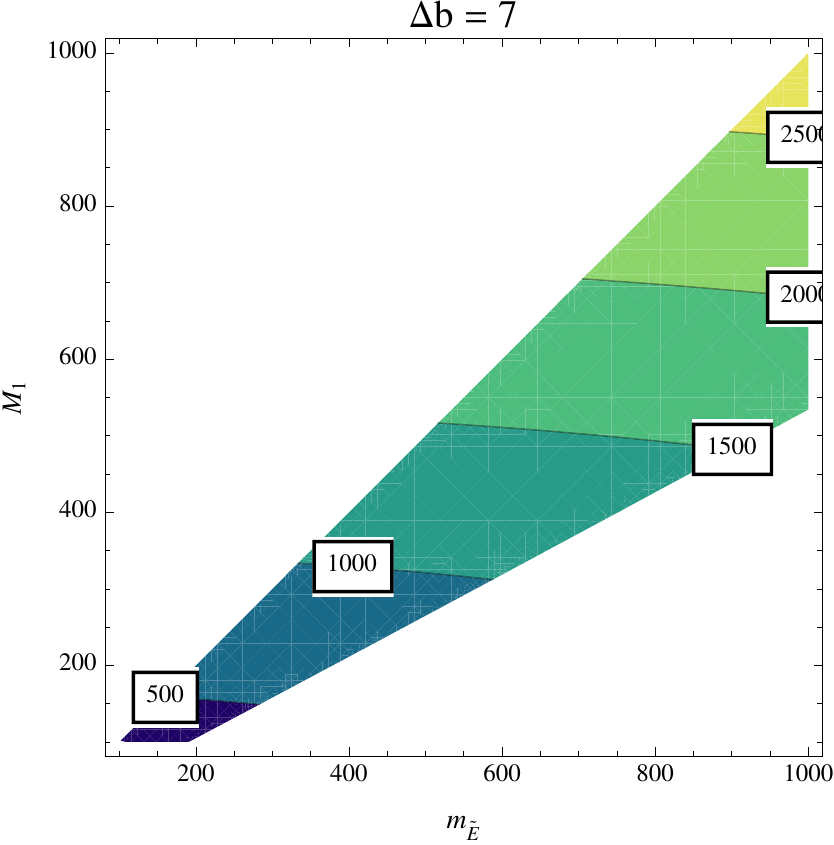}
\hspace{0.2cm}
\includegraphics[height=0.3\textwidth]{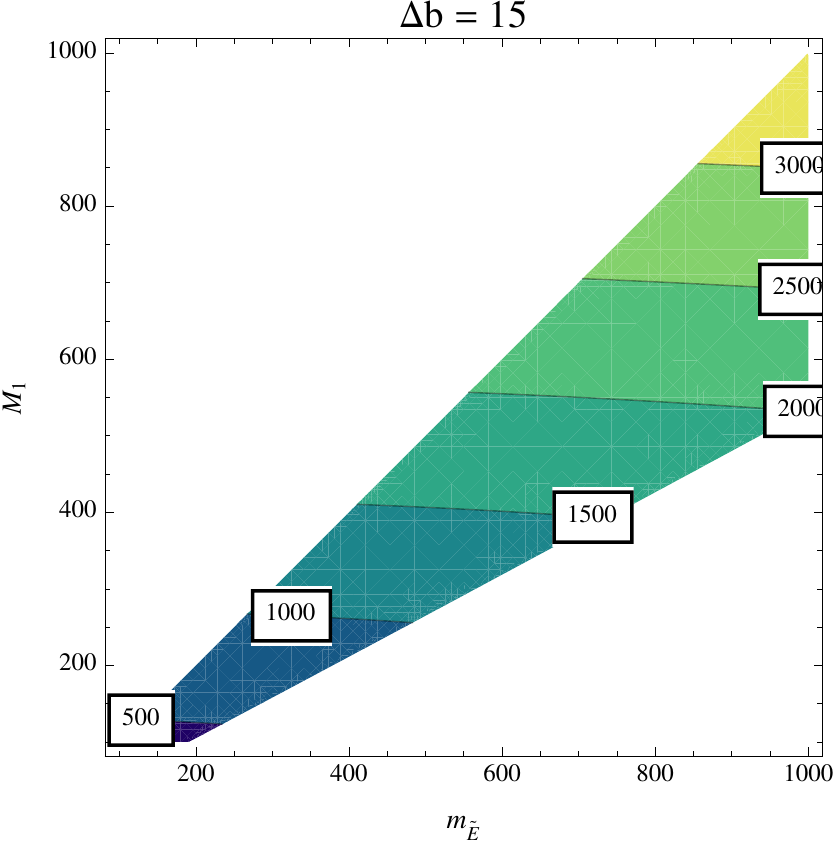}
\caption{Value of the kinematic edge $m_{\ell\ell j}^{\rm max}$ in the
  cascade decay of Fig.~\ref{fig:cascade} on the plane
  $(m_{\tilde{E}},M_1)$ for $\Delta b=0, 7, 15$ at $M_I=10^{13}$ GeV, respectively.}
\label{fig:edges}
\end{figure}

As an example, in Fig.~\ref{fig:edges} we plot contours for
$m_{\ell\ell j}^{\rm max}$ in the plane of the physical masses
$(m_{\tilde{E}},\,M_1\approx m_{{\tilde \chi}^0_1})$ for three
different choices of $\Delta b$ and a fixed $M_I$.  As expected, the
value of the edge changes with the intermediate scale. Again, if it is
possible to independently measure the RH slepton and the neutralino
masses (e.g.~from $m_{\ell\ell}^{\rm max}$ and $m_{\ell j}^{\rm max}$
together with jets and missing $E_T$ distributions) and the edge of
$m_{\ell\ell j}$, this could provide a further way to test
intermediate scale physics.
\begin{figure}[t]
\centering
\includegraphics[width=0.4\textwidth]{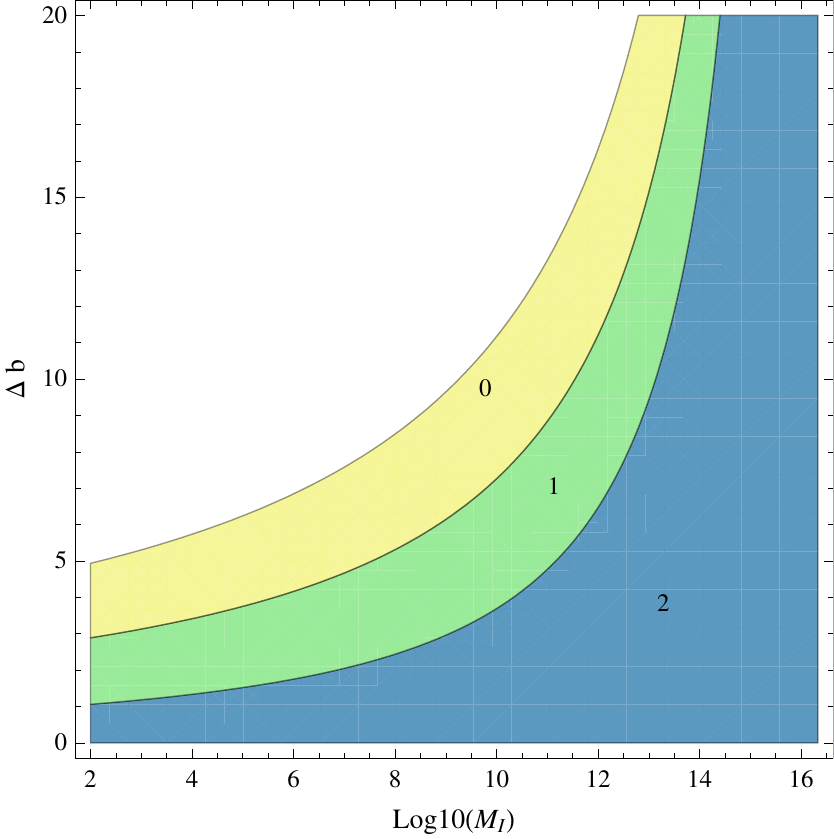}
\caption{Maximum number of edges in the cascade decay of
  Fig.~\ref{fig:cascade} on the plane $(M_I,\Delta b)$.}
\label{fig:maxnumedges}
\end{figure}

As we have mentioned above, depending on the spectrum (in particular whether the sleptons are lighter than the second neutralino) the number of edges can vary. 
Indeed, as we have extensively discussed, the effect of the
intermediate scale is precisely that of increasing the ratio of the
scalar over gaugino masses and thus making the condition of
Eq.~(\ref{eq:cascade}) more difficult to be satisfied. In order to
illustrate this point, we have obtained the maximum number of possible
edges in the $m_{\ell\ell}$ and $m_{\ell j}$ invariant mass distributions
as a function of $(M_I,\Delta b)$. This quantity is independent of the details of the spectrum. The result is shown in Fig.~\ref{fig:maxnumedges}. As we can see, the effect of the 
intermediate scale can be such that the cascade decay of Fig.~\ref{fig:cascade} is never kinematically allowed (in other words $m_{\tilde{\ell}_{L,R}}> m_{{\tilde\chi}^0_2}$ always) 
or just for the lighter sleptons for any choice of sfermion and gaugino masses at high energy. 
It is then clear that, if for instance two clear edges will be observed in di-electron or di-muon distributions, we will be able to exclude a large portion of the $(M_I,\Delta b)$ 
parameter space. 
On the other hand, if no edges are observed at all, intermediate scale physics would remain unconstrained as it is always possible to choose high-energy initial conditions such that
$m_{\tilde{\ell}_{L,R}}> m_{{\tilde\chi}^0_2}$ .

\section{Neutralino Dark Matter}
\label{sec:DM}

\begin{figure}[t]
\centering
\includegraphics[height=0.4\textwidth,angle=-90]{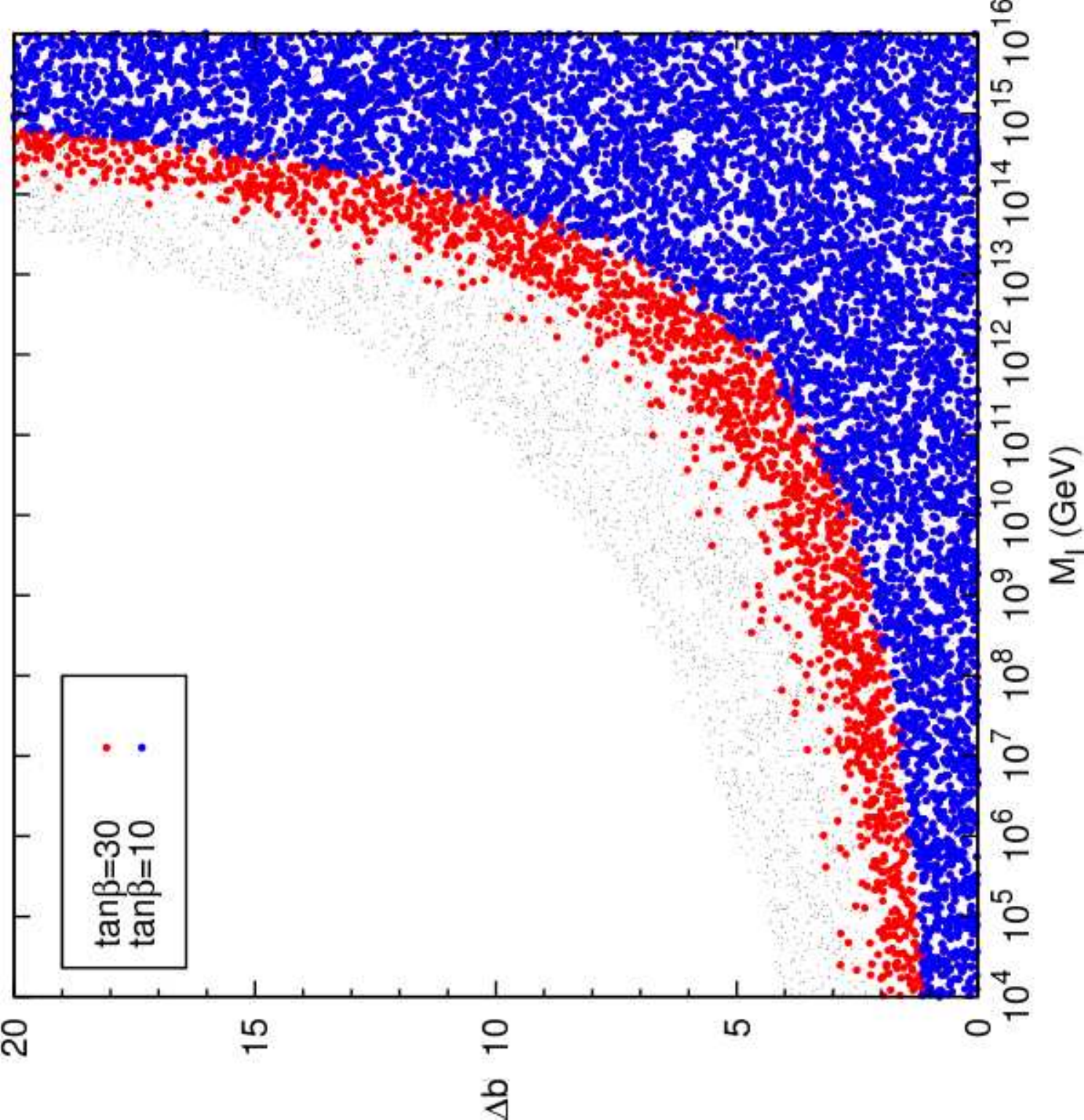}
\hspace{0.5cm}
\includegraphics[height=0.4\textwidth,angle=-90]{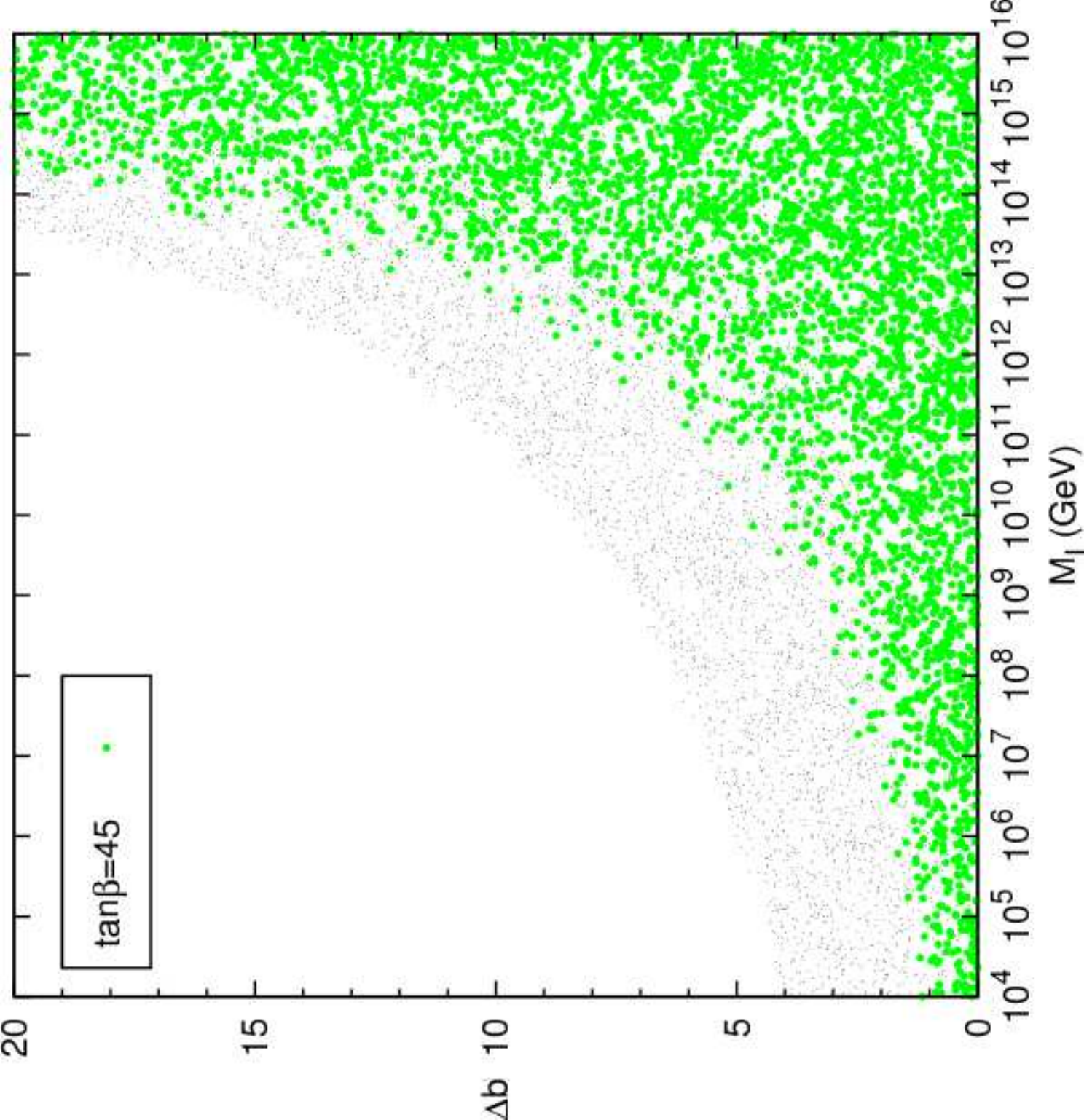}
\caption{Left: region on the $(M_I, \Delta b)$ plane where the correct
  relic density for $\tilde{\chi}^0_1$ is obtained via $\tilde{\tau}$
  coannihilation (blue points corresponds to $\tan \beta=10$, red to
  $\tan \beta=30$). Right: the same for the A-funnel region. The grey
  dots in the background mark the region consistent with the
  perturbativity bounds.}
\label{fig:DM}
\end{figure}

The modification of the SUSY spectrum described in the previous sections can destabilise the
regions of the parameter space where precise relations among the parameters are required in order to fulfill the 
WMAP bound on DM relic abundance. 

It is well known that the lightest neutralino of the MSSM, $\tilde{\chi}^0_1$, typically $\tilde B$-like (i.e.~$m_{\tilde{\chi}^0_1}\simeq M_1$), is overproduced in the early universe, 
unless the neutralino (co)annihilation
cross-section is enhanced by particular conditions. Such conditions define few regions of the parameter space where the WMAP bound is satisfied:
(i) the $\tilde{\tau}$ coannihilation region, where the correct relic density is achieved thanks to 
an efficient $\tilde{\tau}$-$\tilde{\chi}^0_1$ coannihilation, which requires $m_{\tilde{\tau}_1} \approx m_{\tilde{\chi}^0_1}$ \cite{coann};
(ii) the ``focus-point'' region, where the Higgsino-component of $\tilde{\chi}^0_1$ is sizable, i.e.~$\mu \approx M_1$~\cite{focuspoint};
(iii) the A-funnel region, where the neutralino annihilation is enhanced by a resonant s-channel CP-odd Higgs exchange, if $m_{A} \simeq 2\times m_{\tilde{\chi}^0_1}$~\cite{funnel}.

Let us consider first the $\tilde{\tau}$ coannihilation region.  In
the CMSSM, such a region is usually a thin strip which runs along the
border of a wide region of the parameter space excluded because it
gives a $\tilde{\tau}$ as the lightest SUSY particle (LSP)
($m_{\tilde{\tau}_1} < m_{\tilde{\chi}^0_1}$).  As we have seen, the
modification of the running due to the additional fields tends to
increase the scalar masses compared to the gauginos.  The lightest
$\tilde{\tau}$ mass is approximately given (under the conditions
$m^2_{\tilde{\tau}_L}\gg m^2_{\tilde{\tau}_R},~\mu\, m_\tau
\tan\beta$) by:
\begin{equation}
 m^2_{\tilde{\tau}_1} \approx m^2_{\tilde{\tau}_R} - \frac{(\mu\, m_\tau \tan\beta)^2}{m^2_{\tilde{\tau}_L}}\, .
\label{eq:mstau}
\end{equation}
As we have seen, the ratio $m^2_{\tilde{\tau}_R}/M^2_1$ can be strongly increased by the intermediate scale. In particular, it can become larger than one, 
even for vanishing $m^2_{\tilde{\tau}_R}$ at $M_{\rm GUT}$ (see Fig.~\ref{fig:ratio-l}).
As a consequence, the region with a $\tilde{\tau}$ LSP tends to be
reduced and can even disappear.  In fact, even though the second term
of Eq.~(\ref{eq:mstau}) tends to decrease $m^2_{\tilde{\tau}_1}$, the
intermediate scale can easily make the condition $m_{\tilde{\tau}_1} <
m_{\tilde{\chi}^0_1}$ impossible to obtain for any choice of the SUSY
parameters at the GUT scale.  This has been observed in
Ref.~\cite{Calibbi:2006nq} for a qualitatively similar scenario, where
the SU(5) RG evolution from a universality scale above $M_{\rm GUT}$
have been considered.  For the reasons explained above, this is a
general consequence of models with intermediate scales.\footnote{See
  for instance
  Refs.~\cite{Esteves:2010ff,Esteves:2011gk,Esteves:2009qr,Calibbi:2009wk}.}
Clearly the coannihilation strip gets modified as well: it can be
distorted or it can even disappear (in the case of a too large
increase of $m_{\tilde{\tau}_R}/M_1$ such that the condition
$m_{\tilde{\tau}_1} \simeq m_{\tilde{\chi}^0_1}$ cannot be achieved
anymore).  This has been discussed in Ref.~\cite{Calibbi:2007bk} again
in the context of SU(5) RG running.  In Ref.~\cite{Biggio:2010me} the
same effect has been studied in a well-motivated case of
intermediate-scale physics: a type I+III seesaw model, achieved with a
single SU(5) adjoint representation, ${\bf 24}$ (corresponding to
$\Delta b = 5$).  It was shown that, for certain choices of the
parameters, the resulting coannihilation region is bounded from above
(i.e. neutralino DM is only possible in a limited range of the
neutralino mass).

Now let us generalise these observations to generic sets of
intermediate-scale fields.  In the left panel of Fig.~\ref{fig:DM}, we
show the regions on the plane $(M_I,\Delta b)$ where the
coannihilation can take place. The plot was done solving the RGEs
numerically at two loops with CMSSM-like boundary conditions. The blue
points correspond to $\tan\beta=10$, the red ones to $\tan\beta=30$
(the grey dots in the background mark the region consistent with the
perturbativity bounds).  Comparing Fig.~\ref{fig:DM} with
Fig.~\ref{fig:perturbativity}, we see that already values of
$\alpha_U$ as large as $1/15\div 1/10$ are enough to off-set
completely the $\tilde \tau$ coannihilation condition, since
$m_{\tilde{\tau}_1}$ results consistently larger than
$m_{\tilde{\chi}^0_1}$ everywhere in the parameter space.  The effect
can be partially relaxed in the case of very large $\tan\beta$ (and
large A-terms) but intermediate scales configurations corresponding to
$\alpha_U \gtrsim{O}(0.1)$ will still make the coannihilation
region disappear. Dropping the universality assumption
  at $M_{\rm GUT}$, one can still find corners of the parameter space
  where the coannihilation is possible (e.g.~with $m_{H_d}(M_{\rm GUT})
    \gg m_{{\tilde \tau}_R}$ and large $\tan\beta$). 

The A-funnel region can face a similar fate. In fact, the condition  $m_{A} \simeq 2\times m_{\tilde{\chi}^0_1}$ can be made theoretically unaccessible. 
At tree level the CP-odd Higgs mass is approximately
\begin{equation}
 m^2_A \approx m^2_{H_d} - m^2_{H_u},
\end{equation}
As we have seen in section \ref{sec:ewsb}, the ratios $|m^2_{H_{u,d}}|/M_1$ grow with $\alpha_U$, so that $m_A/M_1$ gets increased too. 
The result is depicted in the right panel of Fig.~\ref{fig:DM}, where we show the region of the plane $(M_I,\Delta b)$ where the A-funnel can be obtained for $\tan\beta=45$.
Again, if $\alpha_U$ is too large the condition $m_{A} \simeq 2\times m_{\tilde{\chi}^0_1}$ can be never realised. We notice, however, that the funnel region can be restored by
choosing proper non-universal values for $m_{H_{u,d}}$ at the GUT scale, as $m_{A}$ and $\mu$ then become free parameter. 

Finally, let us comment about the focus point region. As we discussed in section \ref{sec:ewsb}, the ratio $\mu/M_1$ tends to increase as well, however Fig.~\ref{fig:ratio-muM1} shows that
it is always possible to find configurations with $\mu \approx M_1$, such that the Higgsino component of $\tilde{\chi}^0_1$ is sufficiently large to give a sizeable annihilation cross-section.
The focus point region is therefore the only DM branch which is not destabilised by the intermediate scale, if CMSSM-like boundary conditions are assumed. 
Let us remark that this true under our assumption that the new fields do not have large Yukawa couplings
with the MSSM fields. On the contrary, if this occurs, the focus point region is drastically affected and can even disappear \cite{Calibbi:2007bk}.

\section{Proton decay}
\label{sec:pdecay}

In SUSY GUTs proton decay is typically induced by dimension five operators 
generated by the exchange of coloured Higgs triplets. 
Once a mechanism to suppress it is added, the model is safe,
since the contribution of the dimension six operators from 
the gauge bosons associated to the unified gauge group is typically 
below the current experimental bound set by SuperKamiokande: $\tau(p\to e^+ \pi^0)> 1.29\times 10^{34}$~yrs at 90\%
confidence level~\cite{Nishino:2012rv}.
However, when intermediate scale physics is
present, the enhancement of the unified gauge coupling increases the
proton decay rate induced by the GUT gauge bosons to values
close to the current bounds~\cite{Calibbi:2009wk,Hisano:2012wq}. 
This can thus be used to set further
constraints on the intermediate scale.

The partial decay width of the dominant decay mode is given by \cite{Hisano:2012wq}:
\begin{align}
\Gamma (p\rightarrow \pi^0 e^+)
= \frac{\pi}{4}\frac{\alpha_U^2}{M_X^4}\frac{m_p}{f_\pi^2}
 \alpha_{\rm H}^2|1+D+F|^2\biggl(1-\frac{m_\pi^2}{m_p^2}\biggr)^2
 \bigl[
\bigl(A^{(1)}_R\bigr)^2+\bigl(A^{(2)}_R\bigr)^2(1+|V_{ud}|^2)^2
\bigr] \, .
\label{p-decay_width}
\end{align}
Here $\alpha_H=-0.0112$, $D=0.80$ and $F=0.47$ are parameters related to the hadronic
matrix elements and $M_X$ is the mass of the gauge bosons of the
unified gauge group, which we will take equal to $M_{\rm GUT}$ in our
evaluations. $A_R^{(i)}$ are the renormalization
factors given by $A_R^{(i)}=A_L \cdot A_S^{(i)}$ with the long
distance factor $A_L=1.25$ \cite{Hisano:2012wq} and the short distance factors 
\begin{align}
 A_S^{(i)}=&\biggl[
\frac{\alpha_3(M_Z)}{\alpha_3(M_S)}
\biggr]^{-\frac{3\gamma_3}{2b_3^{SM}}}\biggl[
\frac{\alpha_2(M_Z)}{\alpha_2(M_S)}
\biggr]^{-\frac{3\gamma_2}{2b_2^{SM}}}\biggl[
\frac{\alpha_1(M_Z)}{\alpha_1(M_S)}
\biggr]^{-\frac{3\gamma_1^{(i)}}{2b_1^{SM}}}\nonumber \\
&
\biggl[\frac{\alpha_3(M_S)}{\alpha_3(M_I)}
\biggr]^{-\frac{\gamma_3}{b_3^0}}\biggl[
\frac{\alpha_2(M_S)}{\alpha_2(M_I)}
\biggr]^{-\frac{\gamma_2}{b_2^0}}\biggl[
\frac{\alpha_1(M_S)}{\alpha_1(M_I)}
\biggr]^{-\frac{\gamma_1^{(i)}}{b_1^0}}\nonumber\\
&
\biggl[\frac{\alpha_3(M_I)}{\alpha_3(M_{\rm GUT})}
\biggr]^{-\frac{\gamma_3}{b_3^0+\Delta b}}\biggl[
\frac{\alpha_2(M_I)}{\alpha_2(M_{\rm GUT})}
\biggr]^{-\frac{\gamma_2}{b_2^0+\Delta b}}\biggl[
\frac{\alpha_1(M_I)}{\alpha_1(M_{\rm GUT})}
\biggr]^{-\frac{\gamma_1^{(i)}}{b_1^0+\Delta b}}\, ,
\label{eq:pdecay-ren}
\end{align}
with $\gamma_3=\frac{4}{3}$, $\gamma_2=\frac{3}{2}$, $\gamma_1^{(1)}=\frac{11}{30}$, $\gamma_1^{(2)}=\frac{23}{30}$.
The above expressions can be used to estimate the proton life time:
\begin{equation}
 \tau(p\rightarrow \pi^0 e^+)\simeq 2.2\times 10^{36}~{\rm yrs}~\left(\frac{1/25}{\alpha_U} \right)^2 \left(\frac{M_X}{2\times 10^{16}\,{\rm GeV}} \right)^4.
\label{eq:pdecay-appr}
\end{equation}
From this expression we see that the experimental bound $\tau(p\to e^+ \pi^0)> 1.29\times 10^{34}$~yrs is satisfied
unless the unified coupling is very large $\alpha_U \approx 0.5$. 

\begin{figure}[t]
\begin{center}
\includegraphics[width=0.4\textwidth]{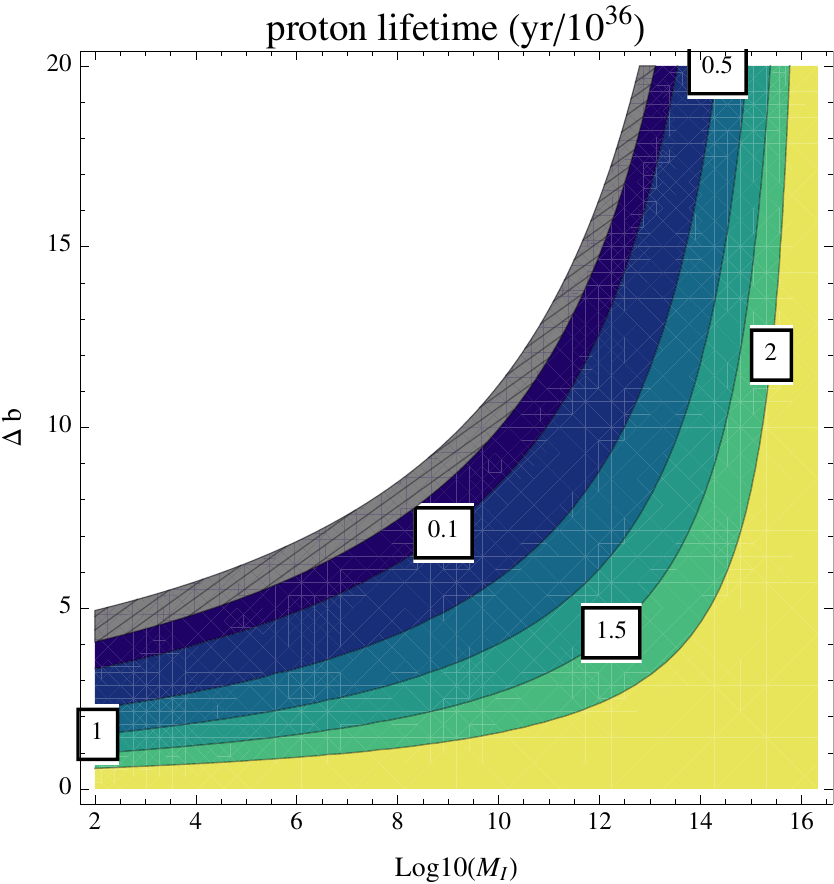}
\caption{Proton lifetime with gauge boson mass $M_X=M_{\rm GUT}\simeq
  2\cdot 10^{16}$~GeV. The
  grey region corresponds to the current exclusion limit.}
\label{fig:p-decay}
\end{center}
\end{figure}

In Fig.~\ref{fig:p-decay} we use Eqs.~(\ref{p-decay_width}, \ref{eq:pdecay-ren})
to plot the proton lifetime on the $M_I$-$\Delta b$
plane. The grey area corresponds to the current exclusion
limit. As expected from the above estimate, we see that the present experimental bound only excludes
a small portion of the parameter space, close to the Landau pole. On the other hand, $\mathcal{O}(1)$ variations
of the GUT vector bosons mass $M_X$ drastically affect the predicted $\tau(p\rightarrow \pi^0 e^+)$, as the estimate
of Eq.~(\ref{eq:pdecay-appr}) shows. Therefore, a proton life time in the reach of the future experimental sensitivity
can be easily obtained for large ranges of the parameters if new fields at intermediate scales are present.

\section{Conclusions}
\label{sec:conclu}

In this paper we have studied the phenomenological consequences of the
presence of new physics at a scale intermediate between the EW and the
GUT scale in a SUSY theory. We have assumed that the new physics
consists only of chiral superfields in complete GUT multiplets, 
such to maintain gauge coupling unification.

The main effect, which drives all the others, is the increment of the
value of the unified gauge coupling. The simple requirement that it
remains perturbative up to the GUT scale is already enough to exclude
a large portion of the parameter space, as we have shown in
Fig.~\ref{fig:perturbativity}. As a consequence of this increase, the
entire low energy spectrum is modified with respect to the MSSM one
and, in particular, the ratio of scalar over gaugino masses is
enhanced. This has interesting consequences both for what concerns
the collider phenomenology and the neutralino dark matter.

We have analysed two main sets of collider observables that can give
hints of the presence of the intermediate scale or, on the contrary,
can be used to constrain it: the mass invariants defined in
Eqs.~(\ref{eq:massinv})-(\ref{eq:CBinv}) and the edges in cascade
decays. By measuring the sparticle masses at the LHC and building the
invariants, one can in principle disentangle if we are in the presence
of a CMSSM-like spectrum or if intermediate scales are present or if
high energy boundary conditions are not universal. The same can be
done if independent measurements of sparticle masses and the position
of the edges in the invariant masses in cascade decays are
available. Still, if this is not the case, we have shown that the
maximum number of edges in these decays can give uncontroversial
information that can be used to constrain the intermediate scale
physics.

On the other hand, the generic increase of the ratio of scalar over
gaugino masses tends to destabilise the regions of the parameters
space where the correct DM relic density is obtained thanks to an
efficient (co)annihilation of the lightest neutralino. This is the
case for the $\tilde{\tau}$-coannihilation region or the A-funnel one:
we have shown that the presence of the intermediate scale can render
impossible to realise the precise relations among the masses of the
involved particles necessary to enhance the (co)annihilation cross
section. On the contrary, in spite of the presence of the intermediate
scale, we found that it is always possible to find regions in the
parameter space where the Higgsino-component of the neutralino is
enough to increase the annihilation cross section and obtain the
correct relic density (``focus-point'' region).

Finally, we have observed that the increment of the unified gauge
coupling can reduce the proton lifetime if the decay is driven by the
GUT gauge bosons. We have shown that, for gauge boson masses equal to
the GUT scale, the actual bound can be used to exclude a small part of
the parameter space.

\section*{Acknowledgements}

SKV acknowledges visits to INFN, Sezione di Padova, and Dipartimento
di Fisica, Univ. of Padova, where this work was initiated and
discussions were possible.  SKV is also supported by DST Ramanujan
Fellowship of Govt of India. CB acknowledge financial support from the
Spanish ministry, project FPA2011-25948.

\appendix
\section{Analytical one-loop formulae}

In this section we collect analytical formulae that can be used to
illustrate the effect of the intermediate scale on the renormalization
group evolution of the MSSM parameters (see Sections
\ref{sec:running}).  Aiming at compact and simply readable
expressions, we consider, besides gauge couplings and gaugino masses,
only first generations sfermion masses, for which Yukawa couplings can
be neglected. Even though it is possible to obtain analytical
solutions of the stop and Higgs masses as well,
they result quite involved, hence we prefer to study these parameters
numerically.

We first write the solution of the (one loop) gauge coupling RGEs in presence of fields at $M_I$ giving a contribution $\Delta b = \sum_i n_i$ (with $n_i$ being the Dynkin index of the SU(5) representation of the $i$-th field) to the $\beta$-function coefficients:
\begin{equation}
 \frac{1}{\alpha_i(\mu)} = \left\{ 
\begin{array}{lr}   
 \frac{1}{\alpha_U}-\frac{b_i}{2\pi} \ln\frac{\mu}{M_{\rm GUT}}\,, &~~ \mu > M_I \\
 \frac{1}{\alpha_U}-\frac{\Delta b}{2\pi} \ln\frac{M_I}{M_{\rm GUT}}
  -\frac{b^0_i}{2\pi} \ln\frac{\mu}{M_{\rm GUT}}\,, &~~ \mu < M_I 
\end{array}
\right.
\label{eq:gauge-sol}
\end{equation}
where $b^0_i$ are the ordinary MSSM coefficients
\begin{equation}
 (b^0_1,b^0_2,b^0_3) = (33/5,1,-3), ~~~ b_i = b_i^0 + \Delta b\,,
\end{equation}
and
\begin{align}
 \frac{1}{\alpha_U} = \frac{1}{\alpha_i (M_S)} - \frac{b^0_i}{2 \pi} \ln \frac{M_{\rm GUT}}{M_S} 
- \frac{\Delta b}{2 \pi} \ln \frac{M_{\rm GUT}}{M_I}\,, \\
\ln \frac{M_{\rm GUT}}{M_S} = \frac{2\pi}{b^0_i-b^0_j}\left( \frac{1}{\alpha_i(M_S)} -\frac{1}{\alpha_j(M_S)}\right)\,.
\end{align}

Analogously for the gaugino masses we have:
\begin{equation}
 M_i(\mu) = \left\{ 
\begin{array}{lr}   
 M_i(M_{\rm GUT})/\left(1-\frac{b_i}{2\pi} \alpha_U \ln\frac{\mu}{M_{\rm GUT}}\right)\,, &~~ \mu > M_I \\
 M_i(M_{\rm GUT})/\left(1-\frac{\Delta b}{2\pi} \alpha_U \ln\frac{M_I}{M_{\rm GUT}}
  -\frac{b^0_i}{2\pi} \alpha_U \ln\frac{\mu}{M_{\rm GUT}}\right)\,, &~~ \mu < M_I 
\end{array}
\right.
\label{eq:gaugino-sol}
\end{equation}
In particular, we have:
\begin{align}
M_i(M_S) = &\frac{M_i(M_{\rm GUT})}{\alpha_U}\,\alpha_i(M_S)  \nonumber \\
= & M_i(M_{\rm GUT}) \left(1 - \frac{b^0_i}{2 \pi} \alpha_i (M_S)\ln \frac{M_{\rm GUT}}{M_S} 
- \frac{\Delta b}{2 \pi} \alpha_i (M_S) \ln \frac{M_{\rm GUT}}{M_I} \right)\,.
\end{align}

For the scalar masses, if we neglect Yukawa and A-term contributions, the solution in general looks like:
\begin{equation}
\label{eq:scalarmasses1}
 m^2_\phi(\mu) = m^2_\phi(M_{\rm GUT}) + K_1(\mu) + K_2(\mu) + K_3(\mu)\,,
\end{equation}
with
\begin{equation}
 K_i(\mu) = C_i \frac{2}{\pi} \int_{\mu}^{M_{\rm GUT}} \alpha_i(\mu^\prime) |M_i(\mu^\prime)|^2 d\mu^\prime \,,
\end{equation}
and for the quadratic Casimirs $C_i$ we have: 
$C_1 = 3/5\, \mathcal{Y}_\phi^2$, with $\mathcal{Y}_\phi$ the hypercharge of $\phi$, $C_2=3/4~(0)$ for SU(2) doublets (singlets),
$C_3=4/3~(0)$ for SU(3) triplets (singlets). 

Using Eqs.~(\ref{eq:gauge-sol}, \ref{eq:gaugino-sol}), we can easily solve the integrals and we get for $\mu < M_I$ 
(in particular $\mu = M_S$):
\begin{align}
\label{eq:scalarmasses2}
 K_i(\mu) = & 2 C_i~M_i^2(M_{\rm GUT})~ 
\left[ \frac{1}{b_i}\left(1 - \frac{1}{\left(1- \frac{b_i}{2\pi}\alpha_U \ln \frac{M_I}{M_{\rm GUT}} \right)^2}\right)
+ \right. \nonumber \\ & \left.
\frac{1}{b^0_i}\left(\frac{1}{\left(1- \frac{b_i}{2\pi}\alpha_U \ln \frac{M_I}{M_{\rm GUT}} \right)^2} -
\frac{1}{\left(1- \frac{\Delta b}{2\pi}\alpha_U \ln \frac{M_I}{M_{\rm GUT}} - \frac{b_i^0}{2\pi}\alpha_U \ln 
\frac{\mu}{M_{\rm GUT}}\right)^2}
\right) \right]\nonumber \\
= & \frac{2 C_i}{b^0_i} \left[\frac{b^0_i}{b_i} M_i^2(M_{\rm GUT}) + \frac{\Delta b}{b_i} M_i^2(M_{I})- M_i^2(\mu)\right]\, .
\end{align}

\section{Effect of two-loop RGEs}
\begin{figure}[!t]
\centering
\includegraphics[height=0.55\textwidth,angle=-90]{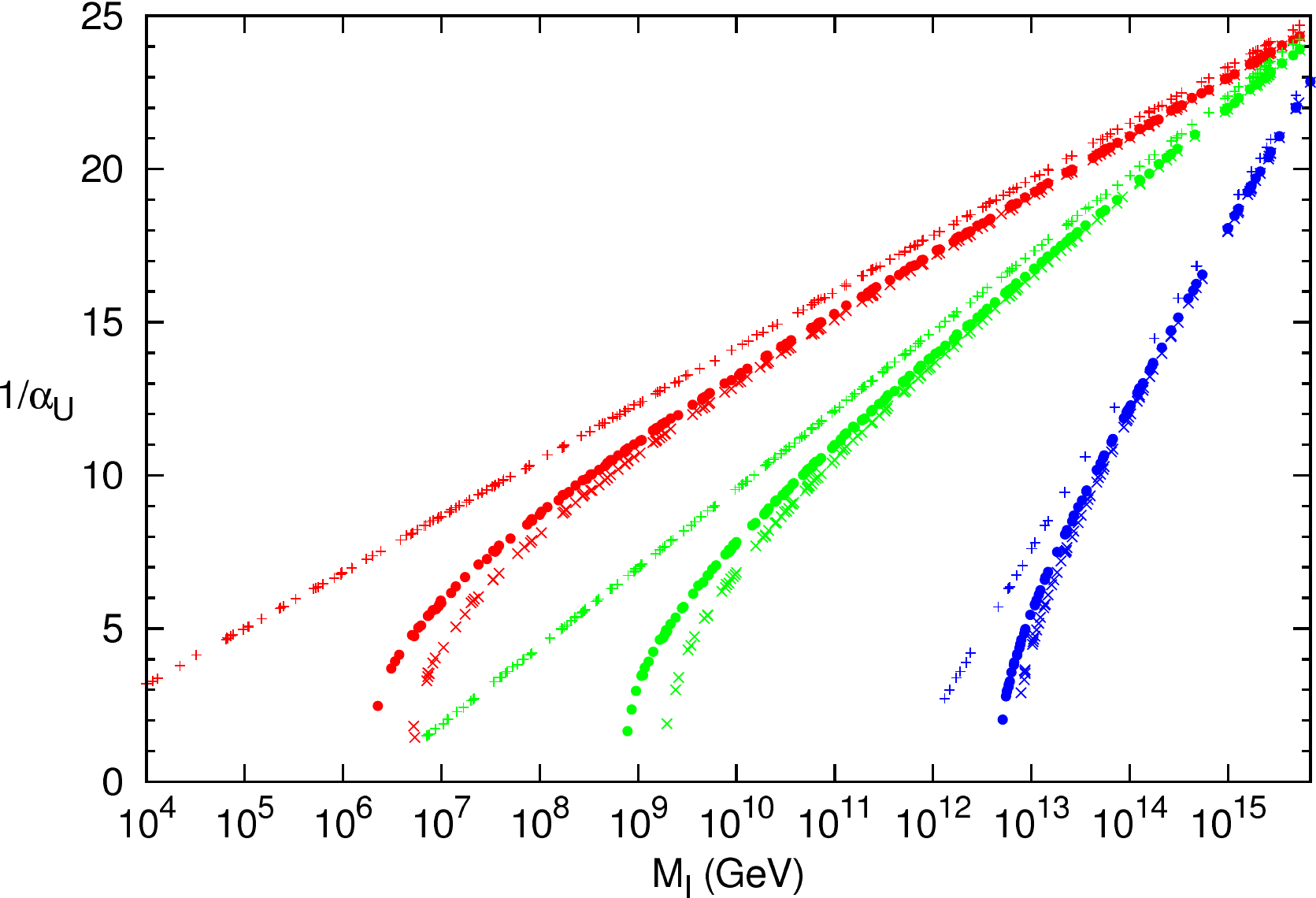}
\caption{Effect of two loop running on $1/\alpha_U$. The red points
  correspond to $\Delta b = 5$, the green ones to $\Delta b = 7$ and
  the blue ones to $\Delta b = 15$. In the three cases, the crosses
  correspond to one loop results, the diagonal crosses correspond to
  the two loop result with $n_{24}=1$ (red), $n_{15}=1$ (green) and
  $n_{24}=3$ (blue) and the dots show the two loop result considering
  only an appropriate number of ${\bf 5+\overline{5}}$.}
\label{fig:summary}
\end{figure}

At two loops, the RGE for the gauge couplings read:
\begin{equation}
 \frac{d}{dt}g_i = \frac{g_i^3}{(4\pi)^2}\,b_i + \frac{g_i^3}{(4\pi)^4}\,\left(\sum_{i} b^{(2)}_{ij}\, g_j^2 - \sum_x C_i^x y_x^2 \right),
\end{equation}
where the two loop $\beta$-function coefficients for the MSSM can be
found, for instance, in Ref.~\cite{martin-vaughn}.  Unlike in the one
loop case, the new contribution depends in principle on the exact
field content at the intermediate scale. For instance, we have:
\begin{align}
&\Delta  b^{(2)} =\\
&\left(
 \begin{array}{ccc}
  \frac{7}{15} n_5 + \frac{23}{5} n_{10} + \frac{181}{15} n_{15}  + \frac{25}{3} n_{24} & 
\frac{9}{5} n_5 + \frac{3}{5} n_{10} + \frac{147}{5} n_{15}  + 15 n_{24} & 
\frac{32}{15} n_5 + \frac{48}{5} n_{10} + \frac{656}{15} n_{15}  + \frac{80}{3} n_{24} \\
  \frac{3}{5} n_5 + \frac{1}{5} n_{10} + \frac{49}{5} n_{15}  + 5 n_{24} &  
7 n_5 + 21 n_{10} + 69 n_{15}  + 45 n_{24} &  
16 n_{10} + 16 n_{15}  + 16 n_{24} \\
  \frac{4}{15} n_5 + \frac{6}{5} n_{10} + \frac{82}{15} n_{15}  + \frac{10}{3} n_{24} &  
6 n_{10} + 6 n_{15}  + 6 n_{24} &  
\frac{34}{3} n_5 + 34 n_{10} + \frac{358}{3} n_{15}  + \frac{230}{3} n_{24} 
\end{array}
\right)\, , \nonumber
\end{align}
where $n_x$ ($x=5,10,15,24$) denotes the number of ${\bf
  5+\overline{5}}$, ${\bf 10+\overline{10}}$, ${\bf 15+\overline{15}}$
and ${\bf 24}$ representations at $M_I$, respectively.
\begin{figure}[!t]
\centering
\includegraphics[height=0.45\textwidth,angle=-90]{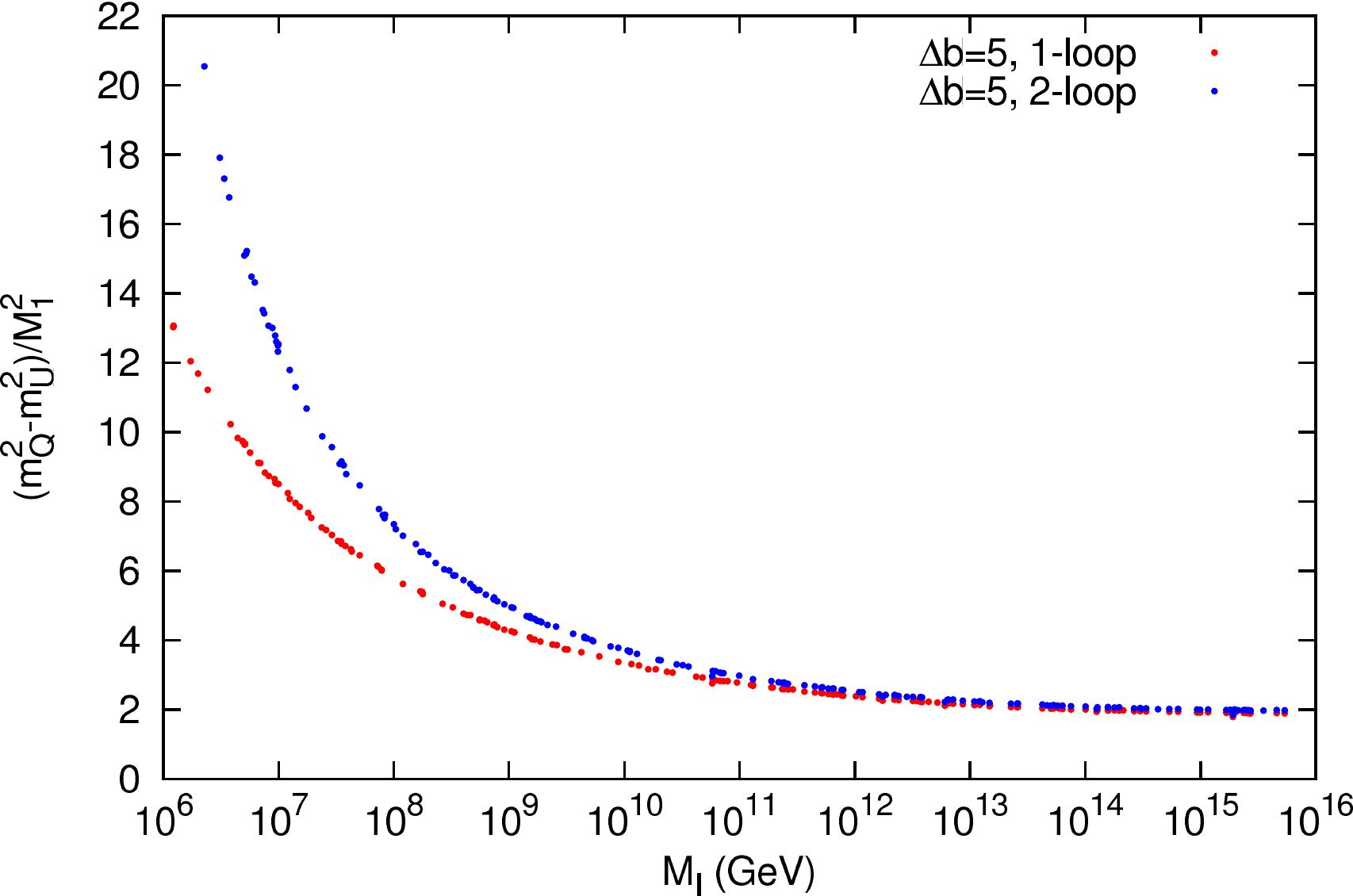}
\includegraphics[height=0.45\textwidth,angle=-90]{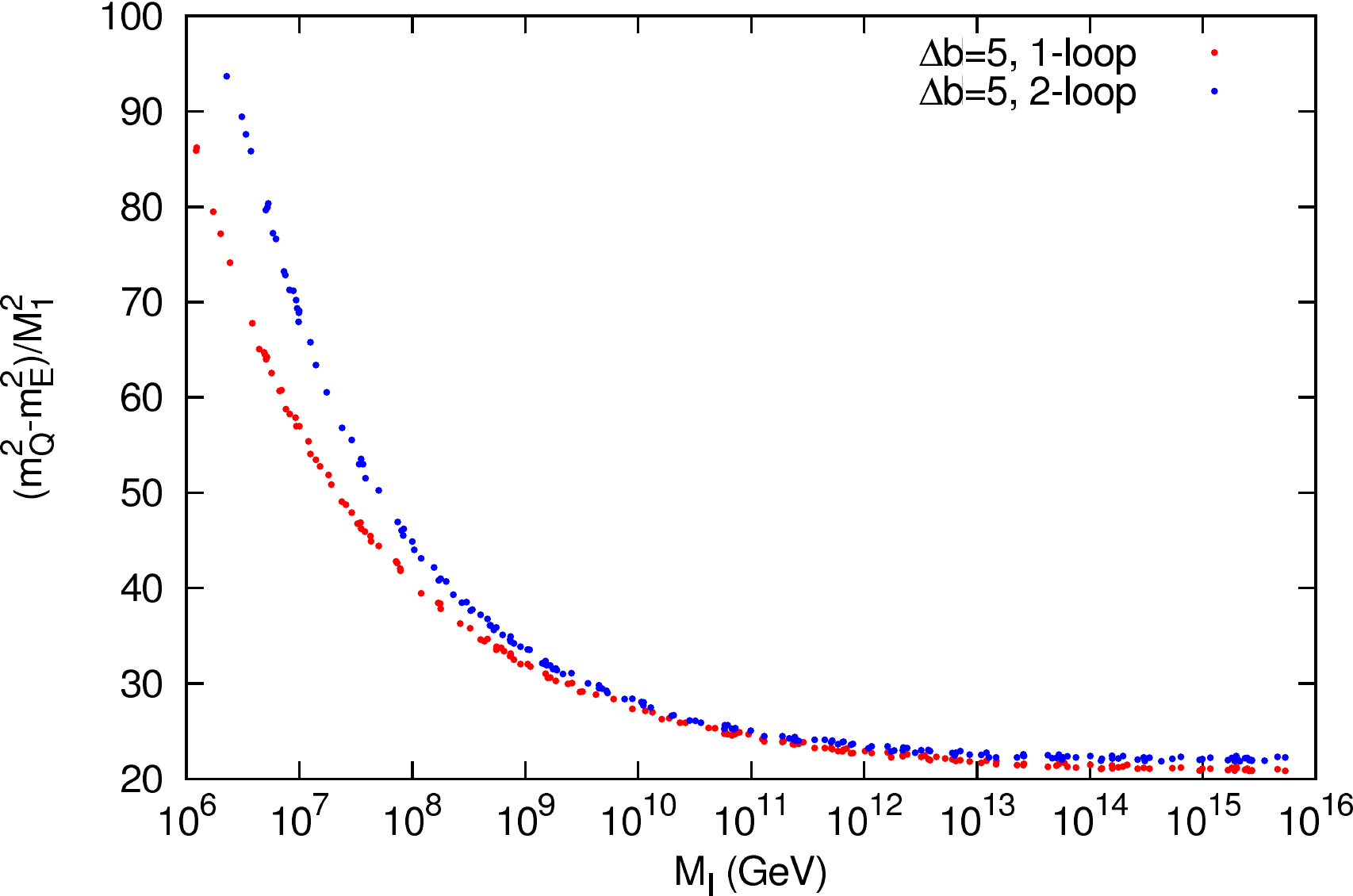}
\includegraphics[height=0.45\textwidth,angle=-90]{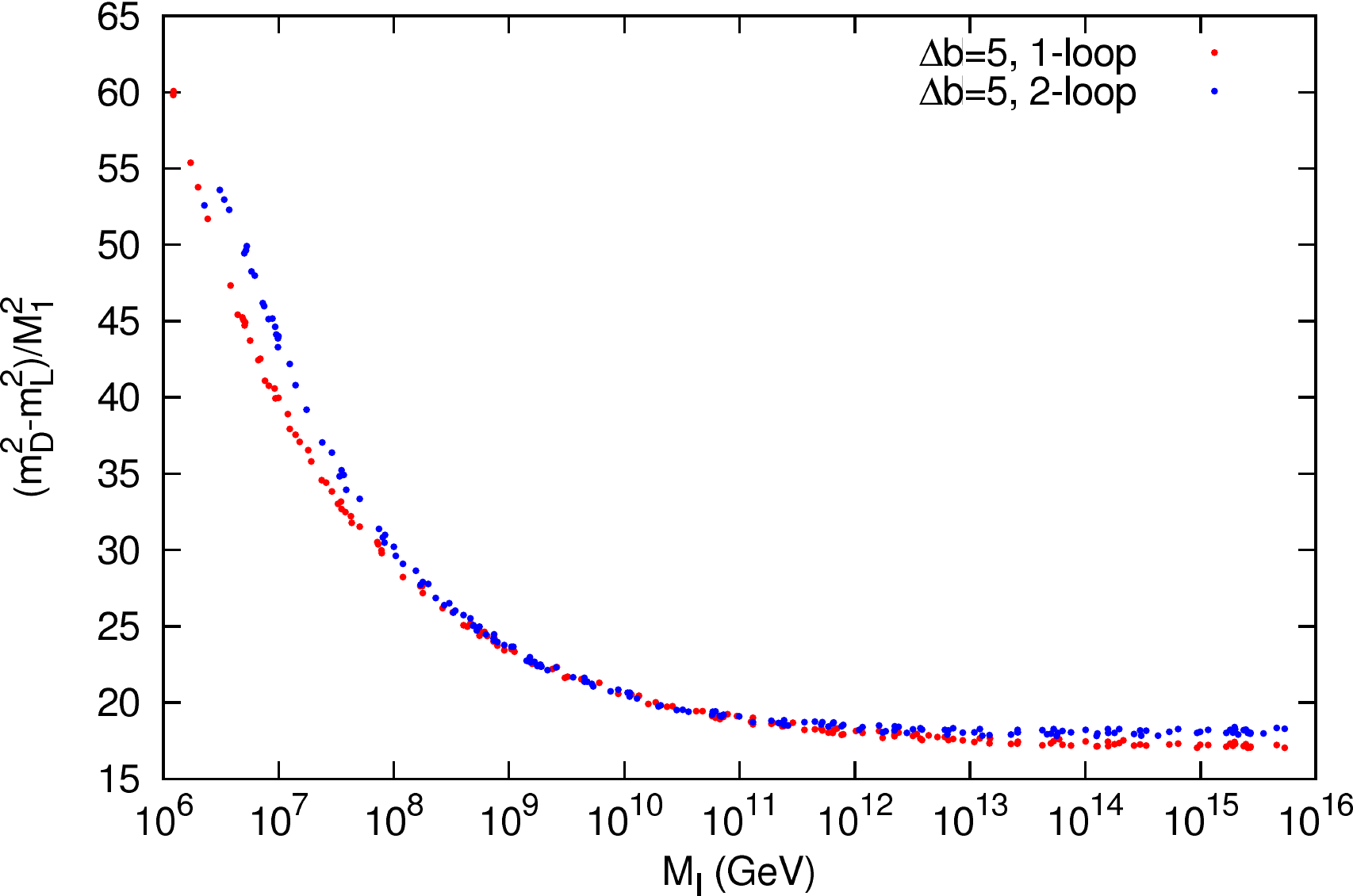}
\caption{Effect of two loop running on the invariants $\Delta^{QU}$,
    $\Delta^{QE}$, $\Delta^{DL}$ for $\Delta b=5$ and $M_S=1$~TeV. The
    red points correspond to one loop RGEs, while the blue ones to two
    loops.}
\label{fig:inv2loop}
\end{figure}
\begin{figure}[!t]
\centering
\includegraphics[height=0.45\textwidth,angle=-90]{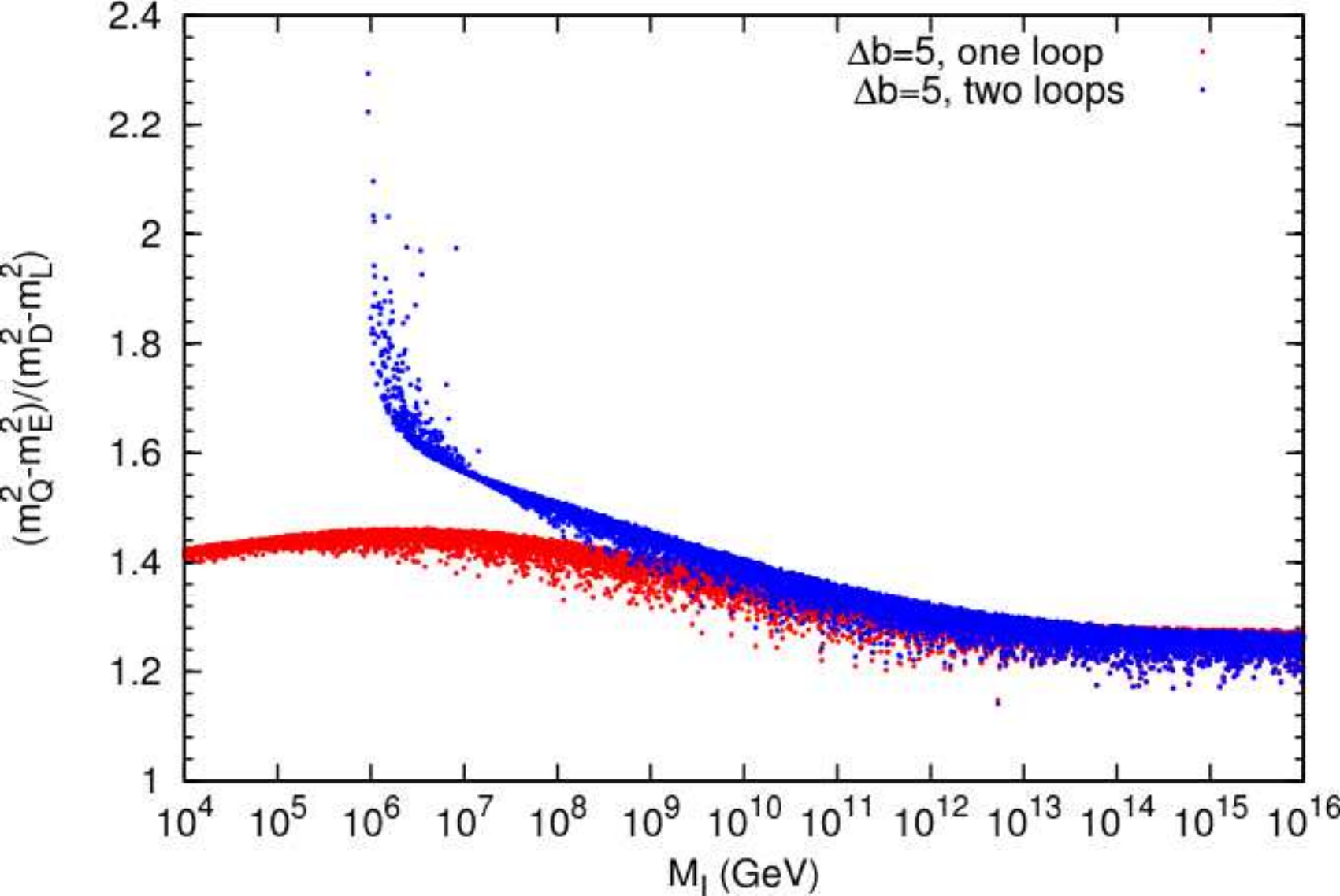}
\includegraphics[height=0.45\textwidth,angle=-90]{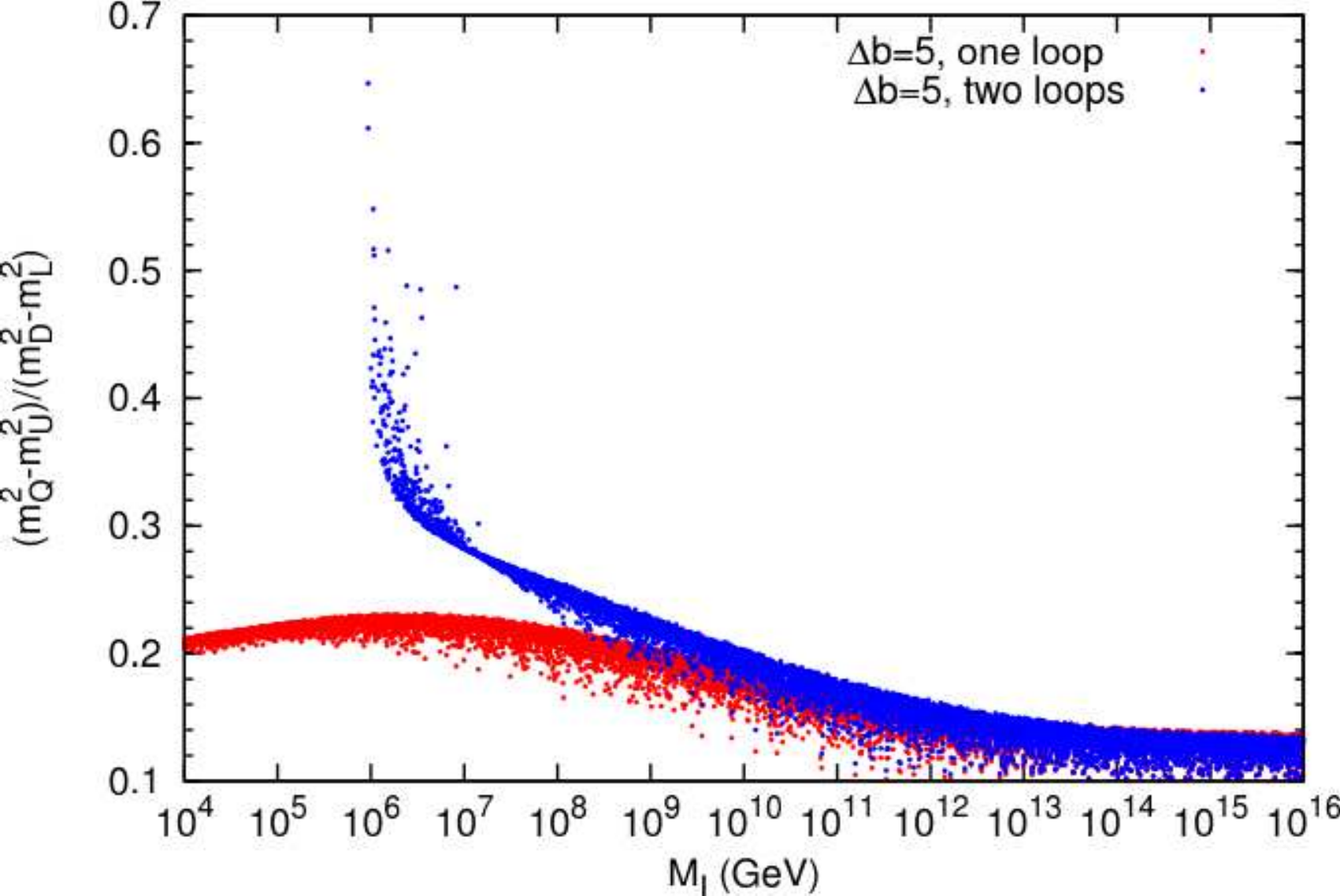}
\caption{Effect of two loop running on the invariants
  $\Delta^{QE}_{DL}$ and $\Delta^{QU}_{DL}$ for $\Delta b=5$ with
  variable $M_S$. The red points correspond to one loop RGEs, while the
  blue ones to two loops.}
\label{fig:inv2loop-bis}
\end{figure}

The effect of two loop RGEs is illustrated in Fig.~\ref{fig:summary},
where the cases $\Delta b = 5$ (red), $\Delta b = 7$ (green) and
$\Delta b = 15$ (blue) are compared. In the three cases, the crosses
correspond to $1/\alpha_U$ vs.~$M_I$ at one loop. The diagonal crosses
correspond to the two loop result with the following field contents:
$n_{24}=1$ (red), $n_{15}=1$ (green) and $n_{24}=3$ (blue).  Finally,
the dots show the two loop result considering only an appropriate
number of $\bf 5+\overline{5}$, i.e.~$n_5 = \Delta b$.  As we can see,
the difference between ``one loop equivalent'' field contents
(e.g.~$1\times {\bf 24}$ and $5\times({\bf 5+\overline{5}})$) is
appreciable only close to the Landau pole. Therefore, the two loop
effects are to a very good approximation independent of the exact
field content and we are going to implement the two loop RGEs taking
always $n_5 = \Delta b$.

In the second panel of Fig.~\ref{fig:perturbativity}, $1/\alpha_U$ is
shown in the $\Delta b -M_I$ plane. We can see that $\alpha_U$
increases with respect to the one loop case. Therefore, we expect all
the effects discussed in the paper to be further enhanced at two
loops.  For illustration, we show in Fig.~\ref{fig:inv2loop} a
comparison between the one loop and the two loops results for the mass
invariants $\Delta_i^{ff^\prime}$ discussed in section~\ref{sec:LHC}
for $\Delta b=5$ and $M_S=1$ TeV.  The two loop correction is mild
(typically $\lesssim 10\%$) and grows close the Landau pole as
expected.  The two loop effect can be much more dramatic in the case
of the quantities $\Delta_{f_3 f_4}^{f_1 f_2}$ defined in
Eq.~(\ref{eq:CBinv}). This is depicted in Fig.~\ref{fig:inv2loop-bis}
again for $\Delta b=5$ and a variation of the SUSY scale $M_S \lesssim
2$ TeV. While $M_S$ has no strong impact on the mass invariants
(contrary to $\Delta_i^{ff^\prime}$), the two loop contribution
can largely increase the effect of the intermediate scale on the
quantities $\Delta_{f_3 f_4}^{f_1 f_2}$.  Therefore the one loop
expressions used for Fig.~\ref{fig:BCinv} just provide a
(conservative) estimate in the regimes where the effect is mild and
two loop RGEs should be taken into account for precise quantitative
studies of this kind of observables.


%
\end{document}